\documentclass[manuscript]{aastex61}
\usepackage{amsmath}
\usepackage[utf8]{inputenc}
\usepackage[T1]{fontenc}
\usepackage{newtxtext, newtxmath}

\usepackage{rotating}  % for sideways figure

\usepackage{savesym}
\savesymbol{tablenum}
\usepackage{siunitx}
\restoresymbol{SIX}{tablenum}
\newcommand{\vect}[1]{\boldsymbol{#1}}

\newcommand{\curly}[1]{\ensuremath{\mathcal{#1}}}
\newcommand{\dd}[2]{\frac{\partial{#1}}{\partial{#2}}}
\newcommand{\diff}[2]{\frac{\d{} #1}{\d #2}}
\newcommand{\lagr}[1]{\frac{\operatorname{D}\! #1}{\operatorname{D}\!t}}

\renewcommand*\d{\mathop{}\!\mathrm{d}}

\newcommand\trad{\tau_\C{rad}}

\newcommand{\C}[1]{\mathrm{#1}}  % constants typeset in upright text
\newcommand\Ro{\mathrm{Ro}}

\begin{document}
	\title{Atmospheric Circulation and Thermal Phase-Curve Offset of Tidally and Non-Tidally Locked Terrestrial Exoplanets}
	\author{James Penn}
	\affil{University of Exeter}
	\email{jp492@exeter.ac.uk}
	\author{Geoffrey K Vallis}
	\affil{University of Exeter}
	\email{g.vallis@exeter.ac.uk}
	\keywords{planets and satellites: tidal locking, atmospheres, exoplanets, terrestrial planets}
	\begin{abstract}
		Using an idealised general circulation model, we investigate the atmospheric circulation of Earth-like terrestrial planets in a variety of orbital configurations.
		We relax the common assumption of the planet being tidally-locked, and look at the role atmospheric dynamics can have in the observed thermal phase curve when the substellar point is non-stationary.
		In slowly rotating planets, a moving forcing can induce strong jets in the upper troposphere, both prograde and retrograde, sensitive to the speed and direction of the diurnal forcing.
		We find that, consistent with previous shallow water model experiments, the thermal phase curve offset is sensitive to the velocity of the substellar point moving across the surface of the planet.
		For a planet with a known orbital period, the results show that the observed hotspot on the planet could be either east or west of the substellar point, depending on whether the planet is tidally-locked or not.
	\end{abstract}
	%!TEX root=./hs_exoplanet.tex

\section{Introduction}\label{sec:introduction}

As computing power and model parameterisations improve, the field of dynamical modelling of exoplanetary atmospheres is becoming increasingly sophisticated.
For many of the more well characterised planets, hot Jupiters in particular we have comprehensive studies of atmospheric circulation \citep[for example]{Komacek2016,Showman2015,Kataria2013}, varying from relatively simple Newtonian relaxation \citep{Menou2009}
to highly parameterised studies including full radiation codes and  disequilibrium chemistry \citep{Cooper2006}.

There has also been an effort to design a thermal relaxation benchmark for inter-comparison of models of tidally locked exoplanets \citep{Heng2011}, drawing on the knowledge gained from this idealisation for Earth modelling, first proposed as a General Circulation Model (GCM) dynamical core inter-comparison benchmark by \cite{Held1994}.

It is likely that many of the exoplanets discovered to date are tidally-locked; due to their close-in orbit of host star the effects of tidal friction lead to a slowing of the planetary rotation rate until it is resonant with orbital rate.
However, this is not true of all planets, as is evident from a quick survey of our own solar system.
Hot Jupiters, very large gas planets, very close to their host star, are subject to strong tidal dissipation forces and so largely expected to be in synchronous orbit; but it is possible that atmospheric winds and internally asynchronous torque forces can force a planet to remain out of a fully tidally locked state \citep{Showman2002}.
There is evidence that this can be the case for rocky planets too; Venus's rotation rate is close to synchronous with it's orbit, yet thermal tides generated in it's thick atmosphere are sufficient to induce a slow retrograde diurnal cycle \citep{Ingersoll1978}.
And recent numerical models have shown that even close-in rocky-exoplanets could be in asynchronous rotation due to the torque induced from an atmospheric thermal tide \citep{Leconte2015}.

The observational detection bias towards large, close-in planets has necessarily driven much of the modelling research effort into tidally-locked exoplanets.
With the assumption of tidal-locking, from the easily observable orbital rate of a planet the more observationally difficult parameter of planetary rotation rate can be inferred; when a planet is tidally locked orbital and rotation rates are equal.

~\cite{Merlis2010} provide a benchmark study for comparing slow and fast rotating Earths in a more complex, moist model that includes the effects of an advected water vapour through both latent heat release and radiative feedback.
In this framework it is shown that the outgoing longwave radiation (OLR) distribution on a tidally-locked Earth is strongly dependent on rotation rate.
The quickly rotating case exhibits a Matsuno-Gill pattern in OLR \citep{Matsuno1966,Gill1980}, with the peak temperatures occurring in trapped Rossby waves west of the substellar point.

The circulation of tidally-locked Earth-like planets has also been considered with detailed land, aquaplanet and cloud resolving parameterisations \citep{Edson2011}.
There it was shown that as rotation rate is reduced from the mean Earth reference, there occurs a sharp regime change from mid-latitude jets to equatorial superrotation at a critical rotation rate, the exact value dependent on the surface parameterisation but it was shown to indeed be a sharp transition.

Equatorial superrotation is an important dynamical feature that is seen in a wide range of exoplanet models, with slow rotation and large scale asymmetric forcing expected of tidally locked hot Jupiters and terrestrial exoplanets orbiting low mass stars \citep{Kopparapu2016}.
It's also observable in the solar system; Venus, a slowly rotating terrestrial planet has a strong superrotating jet at the equator.

Nearly all dynamical models of tidally locked hot Jupiters exhibit a broad superrotating jet transporting heat from the substellar to antistellar point along the equator.
~\cite{Showman2011} use a single shallow-water model to show that the equatorial superrotation can be reproduced in the steady-state Matsuno-Gill framework, when modified to include a mass source term from interaction with an underlying quiescent atmospheric layer.
The superrotating jet is a consequence of large-scale tropical wave forcing.

The equatorial superrotation provides a strong eastward advective force (in this paper, without loss of generality, we will only consider planets rotating in the same directional sense as Earth, and for the sake of clarity use Earth-like nomenclature, so that ``eastwards'' relates to motion prograde with respect to rotation, increasing in longitude).
There is a large temperature forcing gradient between day and night sides, the atmospheric heat content generated at the substellar point is efficiently transported by this jet to the cold night side.
The eastward offset has been a common feature of exoplanets for which sufficient observational evidence has been gathered to generate a thermal phase curve.
For example, the hot Jupiter HD 189733b has an observed eastward offset of 16ºE \citep{Knutson2007} and, more recently, the close-in super-Earth 55 Cancri e was shown to have an eastward thermal phase curve offset of 40ºE \citep{Demory2016}.
The eastward offset of thermal maxima correlates well with the results of dynamical models of tidally-locked hot Jupiters and Earth-like exoplanets.
The spatially large-scale forcing imposed by a tidally locked heating profile forces the atmosphere in the wavenumber 1 and 2 modes at the equator, producing an eddy flux convergence that drives and maintains the superrotational state \citep{Kraucunas2005}.
This advective heat transport is observed in the infrared phase curve of simulated planets, resulting in an eastward offset of the hottest point from the substellar point .

In a previous study \citep{Penn2017}, we used a shallow-water model of the atmosphere to investigate the impact of relaxing the assumption of tidal-locking has on the atmospheric circulation of a planet, and specifically the effect that may be observed remotely through the offset of the thermal phase curve.
We demonstrated that in this model of the first baroclinic mode, the phase curve offset is sensitive to both the orbital and rotation rates of the planet.
When the planet is non-tidally locked the hottest region on the planet --- corresponding to the peak in observed thermal phase curve --- can be offset either to the east or west of the substellar point.
The sign and magnitude of the offset were shown to be dependent on the internal wave speed of the atmosphere.

~\cite{Penn2017} showed that in the shallow water model the offset in thermal phase curve is proportional to the ratio of substellar velocity to internal gravity wave speed of the first baroclinic mode.
When the substellar point is moving faster than wavespeed, in general the hottest point lags behind the substellar point; if you were standing on the surface this would correspond to the hottest time of day occurring in the afternoon.
However, when the diurnal cycle is slower, it was shown that the hottest point could in some cases lead ahead of the substellar point, giving a thermal maxima before the stellar zenith.

In this study we extend the shallow water theory into a three-dimensional, vertically stratified domain and examine the thermodynamic response to a moving stellar forcing.
There have been previous studies modelling non-tidally locked planets.
HD 189733b has been modelled in both synchronous and asynchronous rotation \citep{Showman2009}, the extent of asymmetry between orbital and planetary rotation rates causing a significant shift in the position of the phase curve peak of the thermal emission spectra.
In a similar vein to our shallow water model,~\cite{Rauscher2014} examine the possibility of inferring planetary rotation rate from the phase curve observations of asynchronously rotating hot Jupiters, again studying HD 189733b and HD 209458b they found that it was possible to observe both eastward and westward offsets of the planetary hotspot relative to substellar zenith.
A comprehensive study of hot Jupiters in non-synchronous rotation \citep{Showman2015} demonstrated the relationship between the atmospheric radiative timescale and the diurnal timescale.
It was shown that the dynamics undergoes a regime change when the ratio of these timescales goes through unity, transitioning from equatorial superrotation in slowly rotating / highly irradiated planets, to off-equatorial mid-latitude jets more similar to those observed on Earth in quickly rotating systems where zonal temperature gradients are typically much smaller than the equator-pole difference.

These studies have largely focussed on Hot Jupiters and typically in a parameter space constrained to explore the possible configurations of specific candidate planets.
Here we take a idealised approach to the problem, considering a generalised exoplanet and make a systematic investigation of the parameter space encompassed by differential planetary and orbital rotation rates in an approach similar to other exoplanet studies exploring the parameter space around Earth (\citealp{Merlis2010,Heng2011,Kaspi2015a}).

% section introduction (end)
	%!TEX root=./hs_exoplanet.tex

\section{Model}\label{sec:model}

\subsection{Orbital mechanics}

Given the infinitude of possible orbital configurations, we make a few assumptions to constrain our parameter space to a manageable size.
We are interested in the dynamics in response to a diurnal forcing; specifically around the configuration of tidally-locking and in the regime where the timescale of daily forcing is similar to those of wave propagation and thermal relaxation in the atmosphere.
We assume a circular orbit with zero rotational obliquity --- the planet therefore has no seasonal cycle but can still have a periodic diurnal cycle.
It is beyond the scope of this study to investigate the effect of these parameters, but it is certain that they will affect the climate of an exoplanet and observed characteristics.
\cite{Selsis2013a} model close-in ``super Mercuries'' around low mass stars --- even without the inclusion of a dynamic atmosphere it is shown that the thermal inertia of the surface of non-tidally locked planet on an eccentric orbit leads to complex heat distribution on the surface, and as result large changes in the amplitude and offset of an observed phase curve.
We leave this natural extension of the model for future investigation.

For a planet with orbital rate $\Gamma$, rotation rate $\Omega$, no obliquity and no eccentricity, the diurnal period is given by
\begin{equation} \label{eqn:tsol}
	\C{P_{sol}} = \frac{2 \pi}{\Gamma - \Omega}.
\end{equation}
The length of a stellar day on a planet is thus $\C{T_{sol}} = |\C{P_{sol}}|$, and here we are also concious of the sign of the period.
A negative value indicates that, as on Earth, the substellar point will progress from east-to-west across the surface of the planet, positive values give a prograde progression, as on Venus.

At time $t$, the substellar point is located at longitude
\begin{equation} \label{eqn:sublon_tsol}
	\lambda_0(t) = 2 \pi  \frac{t}{\C{P_{sol}}} = (\Gamma - \Omega)t.
\end{equation}
For a tidally locked planet $\Omega = \Gamma$ and therefore the substellar point remains stationary.
For a non-tidally locked planet the substellar velocity at the equator in the zonal direction is given by the differential of orbital and planetary rotation rates.
Given a planetary radius $a$ (here held constant at Earth's value $a=\SI{6371}{km}$) we can write this as a velocity in a local zonal coordinate $x$,
\begin{equation}
	\diff{x_0}{t} = a (\Gamma - \Omega) \equiv s. \label{eqn:substellar_vel}
\end{equation}

\subsection{Dynamical model}
We model the the shallow atmosphere of an Earth-like planet using the dry, hydrostatic primitive equations, and parametrise the effect of stellar forcing and radiative transfer with a linear relaxation of the temperature towards a predefined reference equilibrium profile.
The equations are, in pressure vertical coordinates,
\begin{align}
	\lagr{\vect u} + \vect{f} \times \vect{u} &= - \nabla_p  \Phi - r\vect u,  \label{eq:hmom} \\
	\dd{\Phi}{\ln p} &= -RT, \\
	\nabla_p \cdot \vect u + \dd{\omega}{p} &= 0, \\
	\lagr{T} &= \frac{\kappa T \omega}{p} + \frac{Q}{c_p}, \label{eqn:thermo}
\end{align}
where the prognostic state variables are the horizontal wind vector $\vect u = (u, v)$, temperature, $T$, and vertical velocity expressed in terms of pressure change $\omega \equiv \mathrm{D}p/\mathrm{D}t$.
The total time, or Lagrangian, $\mathrm{D}/\mathrm{D}t$ operator in pressure coordinates is given by
\begin{equation}
	\lagr{} = (\vect u \cdot \nabla_p) + \omega\dd{}{p},
\end{equation}
where $\nabla_p$ is the horizontal gradient operator along isobaric surfaces.

We solve the equations in spherical coordinates $(r, \lambda, \phi)$: radius, longitude and latitude respectively.
Coriolis force in the traditional approximation is given by $\vect f = \vect{\hat k} 2 \Omega \sin \phi$.
This is a so-called ``shallow'' approximation, as we ignore the effects of Coriolis forces in the vertical; to conserve angular momentum we must also assume that flow is confined to a shallow shell on the surface of the sphere.
The radial coordinate is expressed as $r = a + z$, where $a$ is the radius of the planet; since $a \gg z$ we make the shallow-fluid approximation by replacing $r \to a$, and $\partial r \to \partial z$.
The gravitational force, $g$, assumed to be constant, is absorbed into a height-proxy geopotential $\Phi \equiv gz$.
$R$ is the ideal gas constant for dry air and $c_p$ the specific heat capacity of dry air at constant pressure.  $\kappa = R/c_p$.
The equations are solved in pressure coordinates rather than absolute height, so we also have a prognostic equation of surface pressure at the lower boundary,
\begin{equation}
	\dd{p_s}{t} = -\nabla \cdot \int_0^{p_s} \vect u \d p.
\end{equation}

The equations are forced through a Newtonian heating term
\begin{equation}
	\frac{Q}{c_p} = \frac{T - T_{eq}}{\trad}, \label{eq:Q}
\end{equation}
where $T_{eq}$ is the equilibrium heating profile detailed below, and $\tau_{rad}$ is a characteristic timescale of thermal relaxation of the atmosphere.
The system is damped by linear Rayleigh friction in the momentum equation, parameterised by coefficient $r = r(p)$, the inverse of which can be considered as the timescale of frictional forcing.

We use the open-source Isca framework \citep{Vallis2018}, a recently released fork of the Princeton GFDL FMS modelling suite.
Isca solves the hydrostatic primitive equations (\ref{eq:hmom})--(\ref{eqn:thermo}) on a sphere using a pseudospectral dynamical core in pressure coordinates.
All results are from a T42 resolution (approximately 3º notional grid).
Full source code for the model, and experiments, are available online at \url{https://github.com/execlim/isca}.

\subsection{Newtonian heating profile}\label{sub:newtonian_heating_profile}

The heating profile $T_{eq}$ is calculated from a potential temperature profile $\theta_{eq}$, constructed to approximate the convective-radiative equilibrium state of an Earth-like planet with zero obliquity and eccentricity,
\begin{align}
	\theta_{eq} &= \begin{cases} T_0 - \Delta_h(1- \cos\Theta_s) - \Delta_v \cos\Theta_s \log \frac{p}{p_{ref}} & \cos\Theta_s > 0, \\
								T_{strat} & \cos\Theta_s < 0, \end{cases} \\
	T_{eq} &= \max{
		\left(\theta_{eq} {\left(\frac{p}{p_{ref}}  \right)}^\kappa,
		T_{strat}\right)
		}. \label{eq:teq}
\end{align}
\begin{figure}[tb]
	\centering
	\gridline{
		\fig{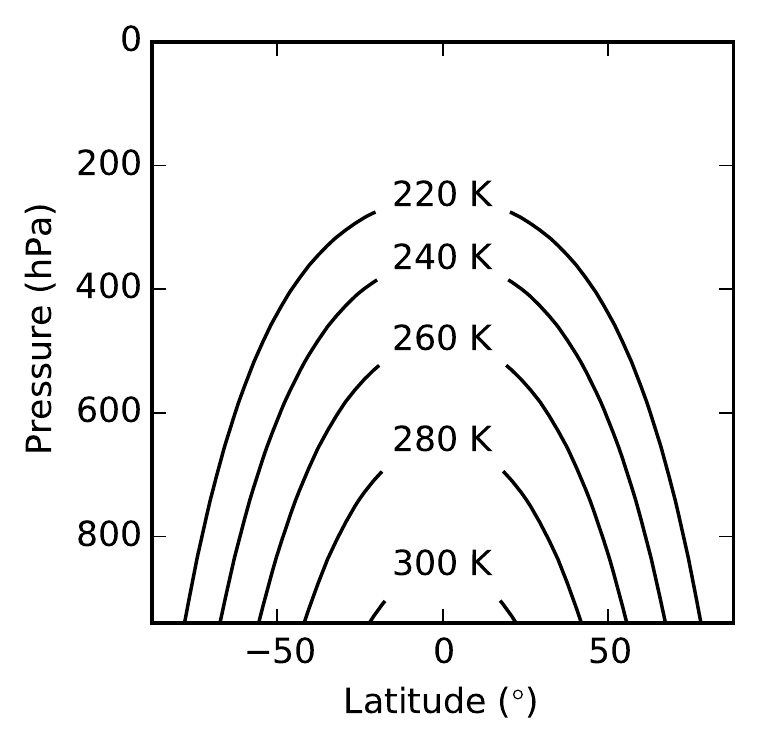}{.35\textwidth}{(a) $T_{eq}$ vertical profile}
		\fig{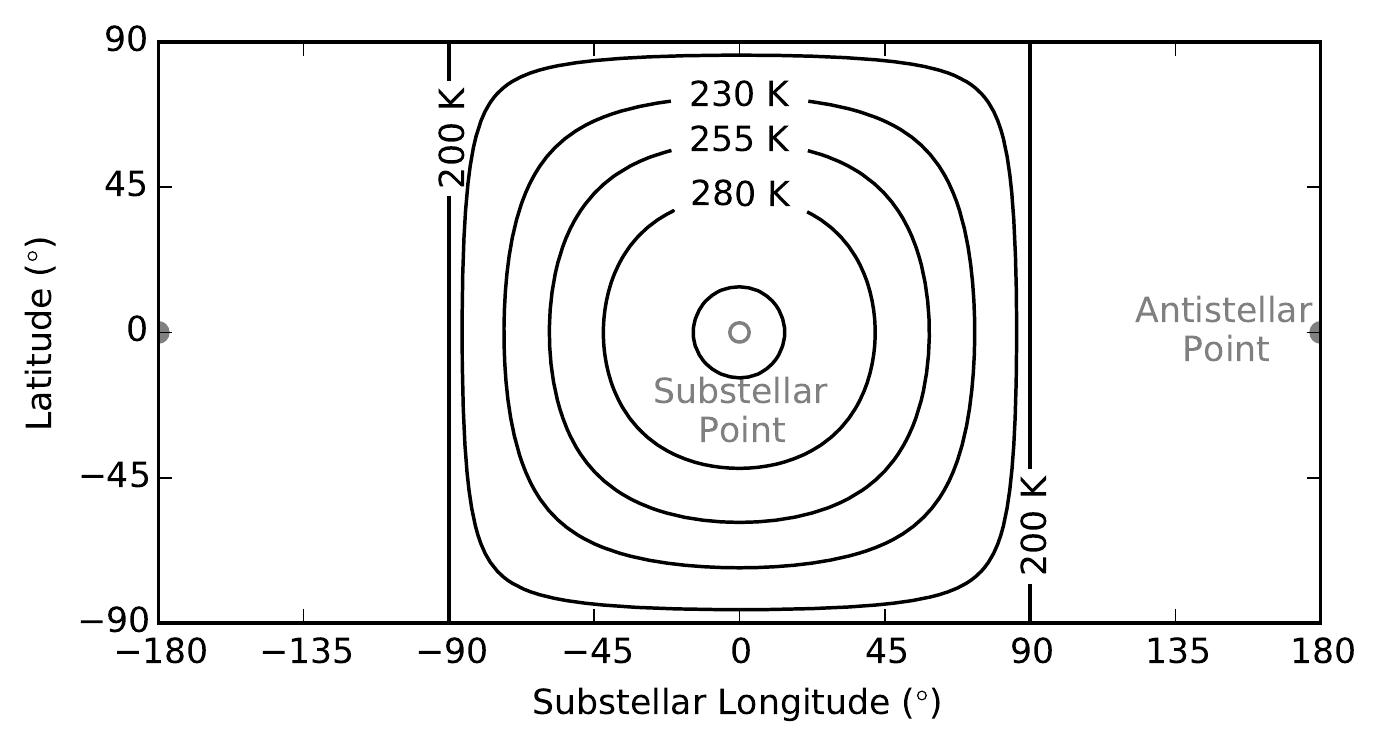}{.63\textwidth}{(b) $T_{eq}$ horizontal profile}
	}
	\caption{Contour plot of Latitude-Pressure (a) and Latitude-Longitude (b) profiles of the relaxation temperature profile $T_{eq}$. The Latitude-Longitude profile is shown with longitude relative to the substellar point.}\label{fig:t_eq}
\end{figure}
Figure~\ref{fig:t_eq} shows the shape of the forcing in height and across the surface of the planet.
This profile assumes an optically thin atmosphere in the shortwave, and no clouds, producing maximal heating at the surface under the substellar zenith, similar to heating from a surface below with zero heat capacity.
Relaxation temperatures are constrained by two temperature extremes: $T_0$ is the heating at the zenith surface, $T_{strat}$ is the background temperature at the stratosphere and dark side of the planet.
The equator-to-pole temperature gradient is $\Delta_h = T_0 - T_{strat}$, and the atmospheric column is marginally statically stable, $\Delta_v = \SI{10}{\per\kelvin}$, such that potential temperature increases gradually throughout the troposphere before becoming very stable in the stratosphere.
The zenith angle of the star, $\Theta_s$ provides the diurnal component of the profile, for a planet with zero obliquity and eccentricity this simplifies to a function of latitude and longitude, and implicitly time, for planets that are not tidally locked
\begin{equation}
	\cos \Theta_s = \cos \phi \cos (\lambda - \lambda_0(t)).
\end{equation}

As it will become the useful reference frame for analysis of numerical results, we define a longitudinal coordinate relative to the substellar point, $\xi = \lambda - \lambda_0$.

As in the formulation of \citep{Held1994}, we apply a timescale of thermal damping that includes a boundary layer
% \begin{equation}
% 	\tau_{rad} = \begin{cases}
% 		\tau_{atm}\left(1- \frac{\sigma - \sigma_b}{1 - \sigma_b} \right) + \tau_{sfc}\left(\frac{\sigma - \sigma_b}{1 - \sigma_b} \right) & \sigma > \sigma_b, \\
% 		\tau_{atm} & \sigma \leq \sigma_b, \\
% 	 \end{cases}
% \end{equation}
\begin{equation}
	\tau_{rad} =
		\tau_{atm} + (\tau_{atm} - \tau_{sfc}) \max\left(\frac{p - p_{BL}}{p_s - p_{BL}}, 0 \right),
\end{equation}
where $p_{BL} = \SI{700}{hPa}$ is an empirically derived pressure height for the top of the boundary layer, but do not scale this damping at the poles as is done in Earth simulations.
Velocity damping is constrained solely to the boundary layer, and decreases linearly with height,
\begin{equation}
	r = \frac{1}{\tau_{fric}}\max\left(\frac{p - p_{BL}}{p_s - p_{BL}}, 0 \right).
\end{equation}
Constants \(\tau_{atm}\), \(\tau_{sfc}\) and \(\tau_{fric}\) have non-trivial impact on the dynamics of the atmosphere, but systematic investigation of this dimension of the parameter space is beyond the scope of this study.
For radiative relaxation we choose values \(\tau_{sfc} = \SI{5}{days}\) and \(\tau_{atm} = \SI{20}{days}\), such that low level heating is more tightly coupled to the forcing profile while at altitude the timescale of relaxation is longer, more akin to the scale of a purely radiative timescale.
Frictional damping, restricted to the planetary boundary layer, is set at \(\tau_{fric} = \SI{1}{day}\).

% subsection newtonian_heating_profile (end)

% section model (end)
	%!TEX root=./hs_exoplanet.tex

\section{Results and Discussion}\label{sec:results}

The numerical model was run with a range of parameter values of $\Omega$ and $s$, as shown in Table~\ref{table:params}.
As we are considering the dynamics of a theoretical exoplanet we do not only restrict the parameter space specific resonant orbital and rotation configurations, instead we vary the substellar velocity independently of the rotation rate.
In the discussion we will use equation (\ref{eqn:substellar_vel}) to address the question of how this may relate to planets trapped in e.g. 3:2 orbital resonances, although this may not be a particularly consistent argument given that we keep orbital eccentricity at zero and thus thermal forcing at a constant, implying a circular orbit.
\begin{table}[t]
	\centering
	\begin{tabular}{l r l}
	\hline
	Rotation rate, \(\Omega\) & 1, 3, 10, 30, 100, 300, 1000 & \(\times 10^{-7}\) \si{\per\second} \\ \hline
	Substellar velocity, \(s\) & -200, -100, -50, -25, -10, -5, 0, 5, 10, 25, 50, 100, 200  & \si{\meter\per\second} \\
	\hline
\end{tabular}
	\caption{Parameter values of rotation rate and substellar velocity.}\label{table:params}
\end{table}

For the sake of clarity, only a representative subset of the parameter space listed in Table~\ref{table:params} are plotted in the following figures, but the complete span of values are considered in the discussion that follows.

\subsection{Dynamics}\label{sub:description_of_zonal_winds}
The atmospheric dynamics are dependent on both the rotation rate and substellar velocity.
The effects of rotation on tidally-locked planets \citep[for example]{Noda2017} and axis-symmetrically heated aquaplanets \citep[for example]{Kaspi2015a} have been well studied in recent years; we build upon these results and will not linger too long on discussing the tidally locked cases in isolation, instead focussing on the inclusion of a moving forcing.

We define the planetary Rossby number,
\begin{equation}
	\Ro = \frac{U}{\Omega a},
\end{equation}
where $U$ is a characteristic zonal velocity as non-dimensional number for characterising the influence of rotation on the large-scale dynamics.
When $\Ro \ll 1$, as is the case on Earth, rotation will constrain atmospheric flow and the dynamical balance of forces, especially away from the equator, will be geostrophic.
At large Rossby numbers rotation plays a much smaller role, pressure gradients largely being balanced by frictional forces.
We will show however, the addition of a moving heat source in both regimes can alter the large-scale structure of the flow.

\begin{figure}[tb]
	\centering
	\includegraphics[width=\textwidth]{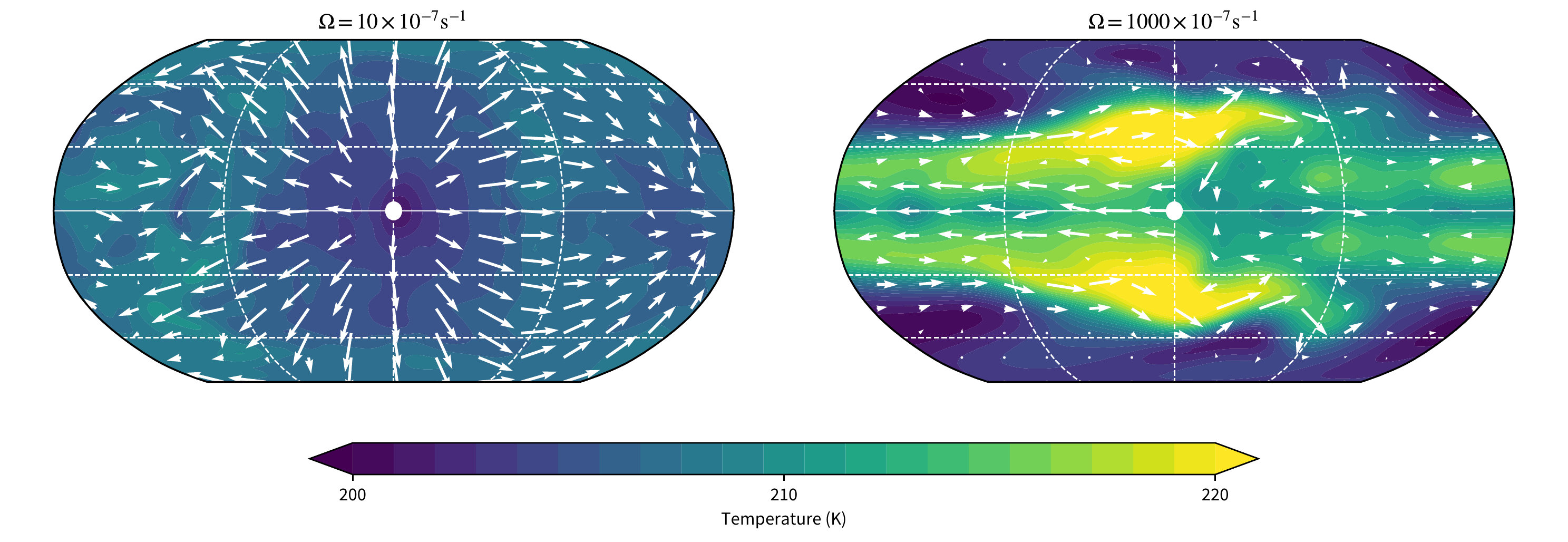}
	\caption{Snapshot of temperature and wind field at the \SI{406}{hPa} level for tidally locked planets rotating slowly (left) and quickly (right).
	The substellar point is in the middle of the domain, shown as a white spot. Compare to Figure~\ref{fig:snapshot_s5} where the substellar point is moving slowly eastward.  The substellar point is in the centre, shown by a white spot.}
	\label{fig:snapshot_s0}
\end{figure}

\begin{figure}[tb]
	\centering
	\includegraphics[width=\textwidth]{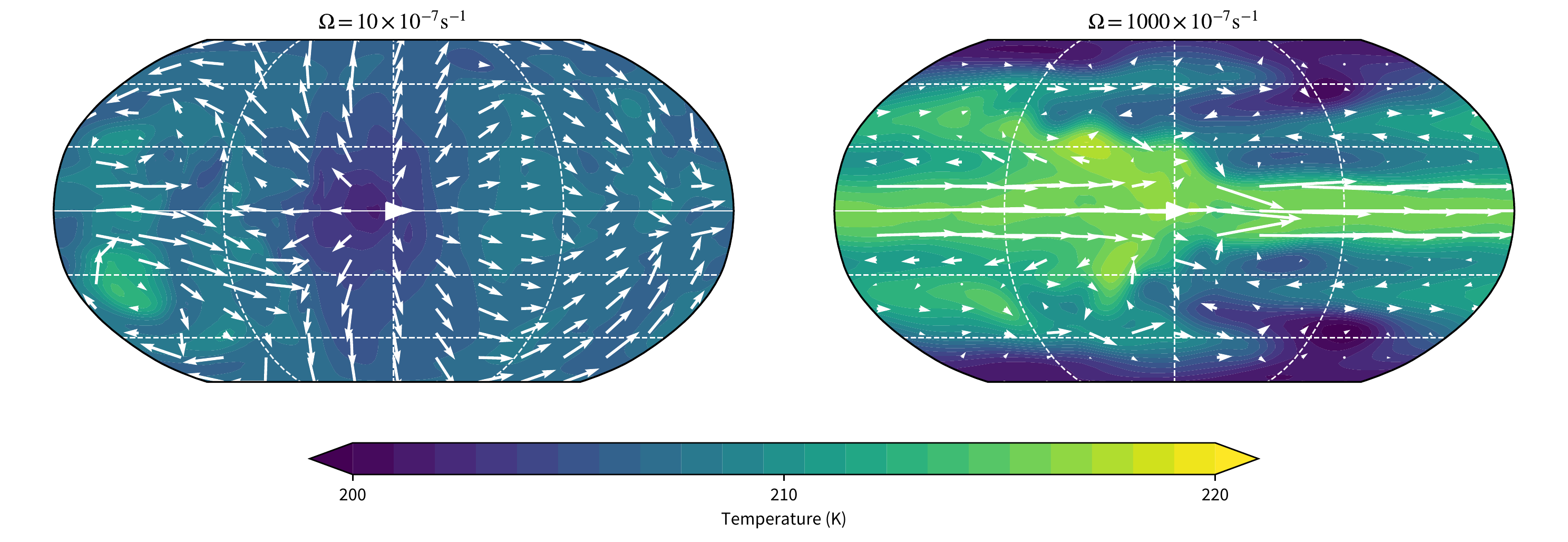}
	\caption{Snapshot of temperature and wind field at the \SI{406}{hPa} level for non-tidally locked planets rotating slowly (left) and quickly (right).  The substellar point is moving eastward at \SI{5}{\meter\per\second}, the substellar point and direction denoted with a white arrow. Compare to Figure~\ref{fig:snapshot_s0} where the substellar point is fixed.}
	\label{fig:snapshot_s5}
\end{figure}

Figures~\ref{fig:snapshot_s0}~and~\ref{fig:snapshot_s5} show snapshots of temperature and wind in the mid-troposphere of slowly rotating ($\Ro\sim 10$) and quickly rotating ($\Ro\sim 0.1$) planets in the left and right-hand panels respectively, after the simulations have been allowed sufficient time to reach a statistical steady-state.

Consistent with the first-baroclinic mode theory of the previously presented shallow water model, at slow rotation rates with a slowly moving (or stationary) diurnal cycle, upper level flow is divergent from the heated hemisphere.
To first order, the response to heating is a overturning circulation spanning the entire planet, updraft at the substellar forcing inducing large scale convergence at the surface and divergence aloft.
The weak Coriolis parameter lends little dynamical asymmetry between the meridional and zonal directions, and thus a thermally direct overturning circulation extends equator-to-pole, day-to-night, air rising near the substellar point and eventually falling at the pole or night side, producing adiabatic heat transport.
The largely horizontally-uniform temperature on the slowly rotating cases has been well explained by the weak temperature gradient (WTG) approximation \citep{Sobel2001,Mills2013}, the small Coriolis parameter means that the dominant balance in the dynamical equations is convective, between heating and vertical advection of potential temperature.

%TODO: [IGNORE] move this section to discussion?
For quickly rotating planets circulation is constrained by the Coriolis force and the response is qualitatively different between a stationary and moving forcing.
With a tidally locked stationary configuration the flow temperature and wind fields display a global Matsuno-Gill pattern of trapped Rossby wave lobes mid-latitudes west of the centre of the forcing, with a small circulation around them and a strong westward flow along the equator, west of the substellar point.
When even a small prograde velocity is applied to the forcing, the circulation changes dramatically, instead inducing a eastward superrotating jet along the equator, with eddies extending towards the poles.

\begin{figure}[tb]
	\centering
	\includegraphics[width=\textwidth]{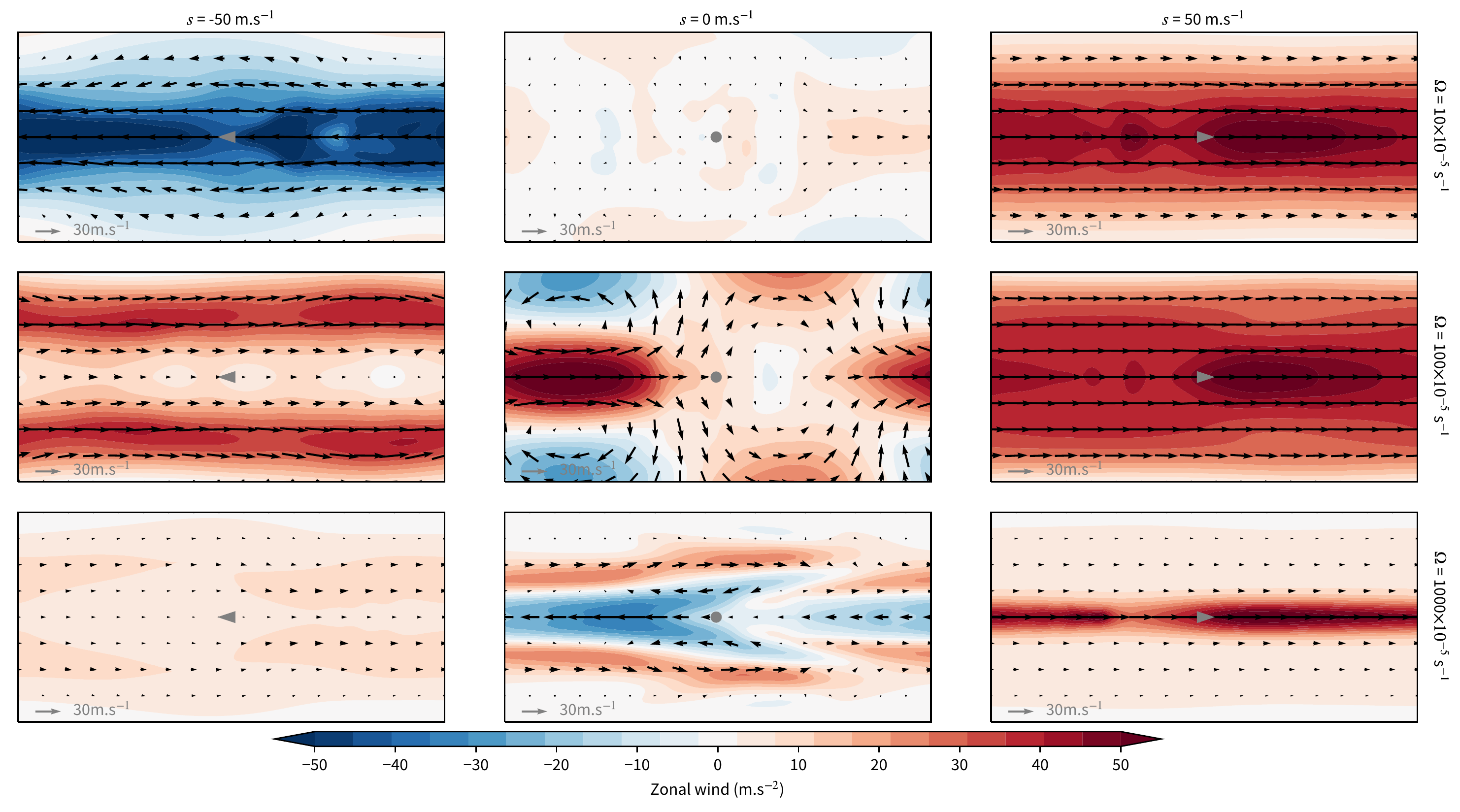}
	\caption{Jet structure in the upper troposphere (p = \SI{249}{hPa}) for increasing rotation rates (rows) and substellar velocities (columns).  Vectors indicate the wind, keyed with \SI{30}{\metre\per\second} in the lower left corner.  Coloured contours show the zonal component of the mean flow only.  The position and direction of motion of the substellar point is indicated with a grey arrowhead.  Time averaged over 50 days after a 500 day equilibrium spin-up period.}\label{fig:zonal_wind_319hPa}
\end{figure}

Figure~\ref{fig:zonal_wind_319hPa} shows the temporal-mean upper-tropospheric flow for a range of retrograde and prograde substellar velocities at increasing planetary rotation rate.
As the substellar velocity increases, increasing strength zonal jets are induced in the tropics, producing strong super rotation when the progression of substellar point is prograde with respect to rotation.
The superrotating jet persists for increasing rotation rate becoming more equatorially constrained as the planetary Rossby number decreases.

A retrograde moving forcing, as on Earth (although on Earth the velocity of substellar point along the equator is $\sim -\SI{450}{\metre\per\second}$, putting in a regime well off to the far left of the parameter values studied here), appears to inhibit the formation of superrotating equatorial jet.
Furthermore, a subrotating equatorial flow becomes a robust feature of the atmosphere of slowly rotating planets with a retrograde forcing.

A westward superrotating jet is generated from a substellar point moving westward at \SI{50}{\meter\per\second} (upper left of figure~\ref{fig:zonal_wind_319hPa})
As the substellar point moves beyond a critical velocity, between $\pm$ \SIrange{50}{100}{\meter\per\second}, slow-rotator circulation transitions again, the upper level zonal flow weakening and eventually turning off.
In these cases, the speed of substellar point means that the diurnal timescale,
$\tau_{diurnal} = 2\pi a / s$,
the time taken for the substellar point to make a complete revolution of the planet, is of the equivalent order as thermal relaxation $\tau_{rad}$.

In the fast rotating cases (lower half of figure~\ref{fig:zonal_wind_319hPa}), circulation is dominated by geostrophic forces.
The flow becomes latitudinally constrained, but the latitudinal cross-section structure is determined by the direction and speed of the moving forcing.
In the fastest rotation cases we observe multiple eddy-driven jets being formed, as expected from quasi-geostrophic theory and as seen in the rotation-rate parameter studies of \cite{Kaspi2015a}.
In the intermediate rotation case $\Omega=\SI{1e-5}{\per\second}$, the tidally-locked and prograde moving substellar point configurations exhibit a single equatorial superrotating jet.
When the substellar point is moved sufficiently quickly retrograde to rotation, $s \simeq\SI{50}{\metre\per\second}$, the equatorial jet is inhibited and instead two mid-latitude westerly jets are formed.

\begin{figure}[tb]
	\centering
	\includegraphics[width=\textwidth]{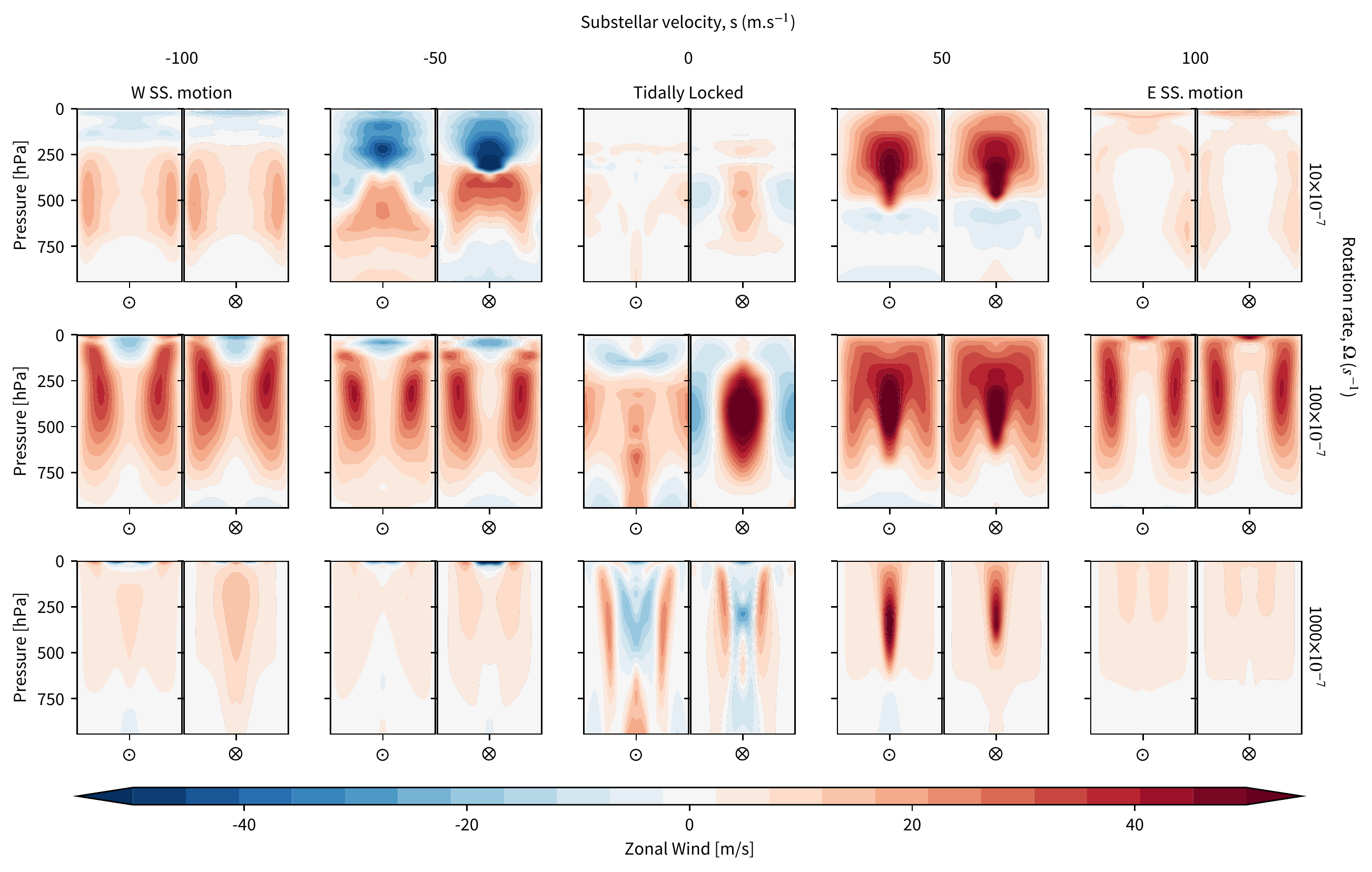}
	\caption{Time averaged zonal-wind latitude-pressure profile at the substellar (\( \odot \)) and anti-stellar (\( \otimes \)) points.
	The abscissa in each pane is latitude from -90º to 90º with the equator denoted by a tick-mark.}\label{fig:ss_as_zonal_wind}
\end{figure}

\begin{figure}[tb]
	\centering
	\includegraphics[width=\textwidth]{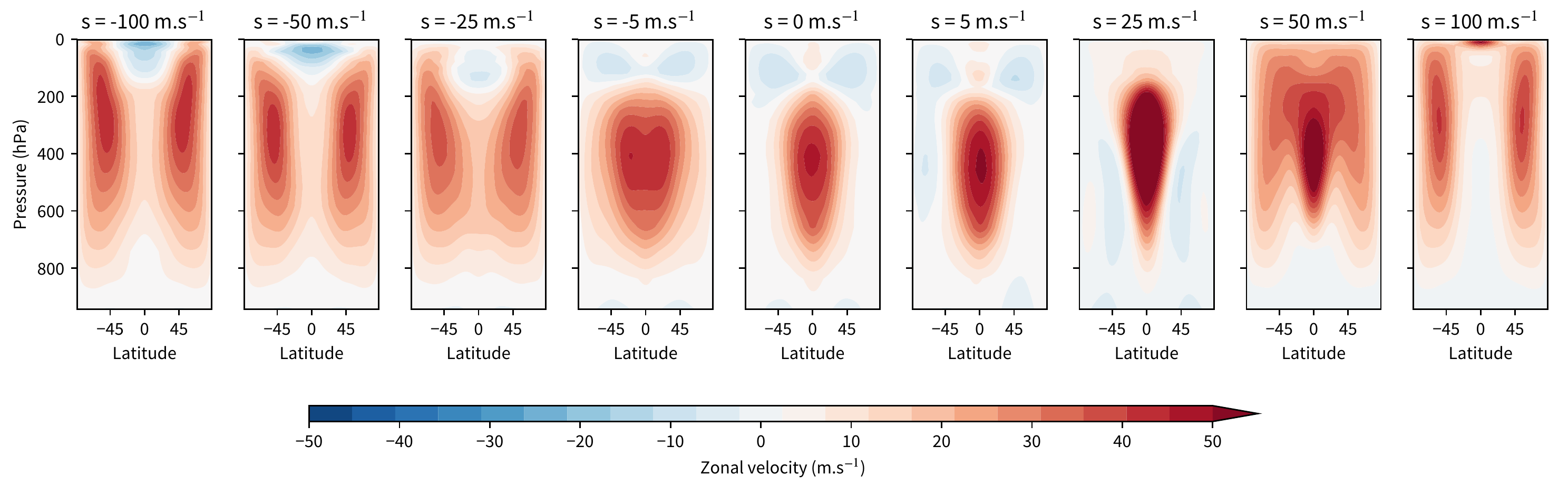}
	\caption{Zonal-mean zonal-wind with varying substellar motion, with rotation rate $\Omega=\SI{100e-7}{\per\second}$.  From left-to-right goes from fastest retrograde substellar velocity, through tidally locked, to fastest prograde motion.}\label{fig:zonal_jets}
\end{figure}

\begin{figure}[tb]
	\centering
	\includegraphics[width=0.6\textwidth]{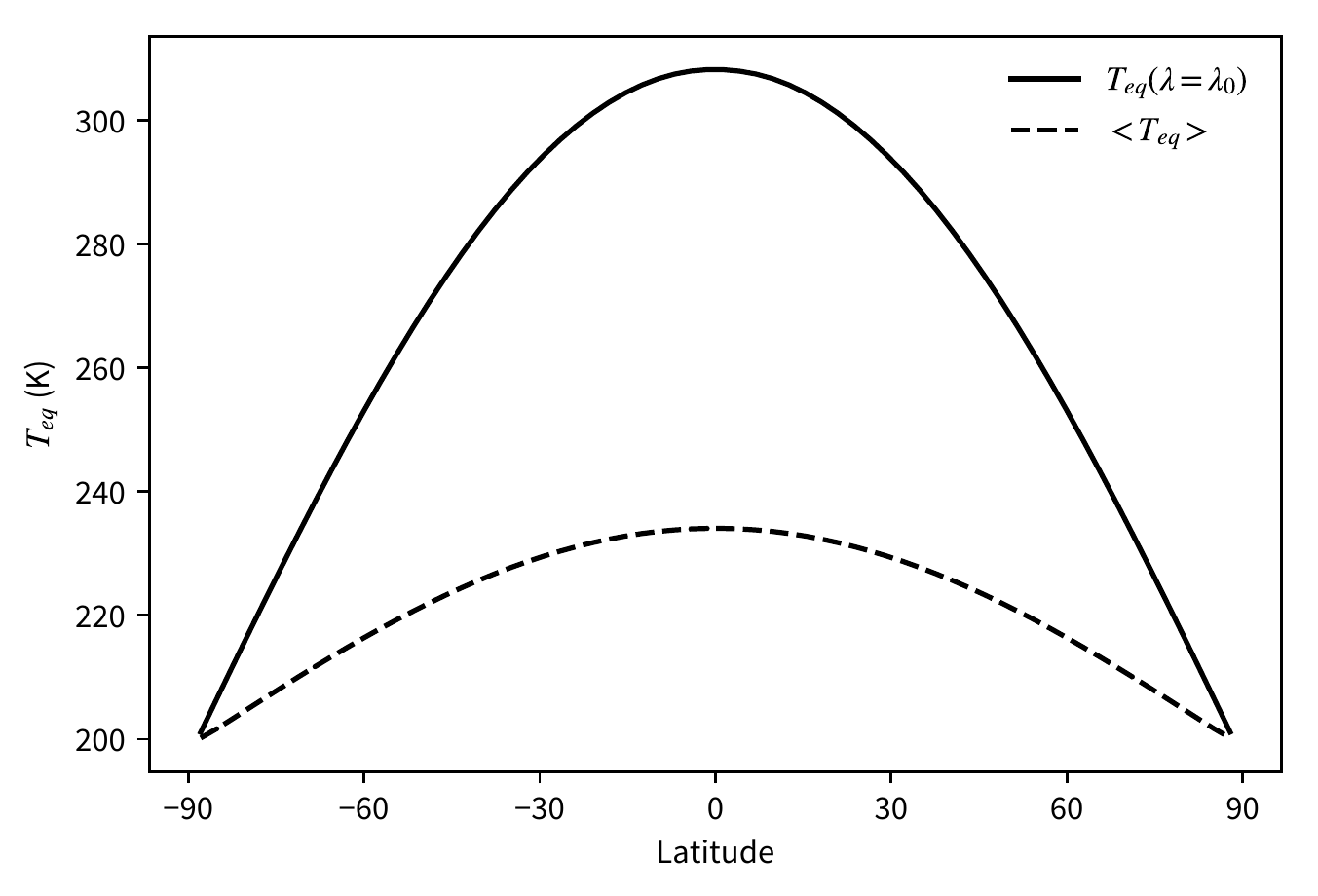}
	\caption{Surface equilibrium temperature gradients at the substellar point (solid) and the zonal mean, $<T_{eq}>$, (dashed).  The much reduced $\Delta_{\mathrm{eq-pole}} T_{eq}$ in the mean versus substellar point impacts the dynamics of planets with very quickly moving diurnal cycles.}
	\label{fig:mean_teq}
\end{figure}

The effect of the retrograde moving forcing on the superrotating jet can be more readily observed in the vertical structure of the atmosphere in both the day and night hemispheres (Figure~\ref{fig:ss_as_zonal_wind}).
When the Rossby deformation radius is small enough to drive zonal flow, but not dominating in the dynamics (the rows labelled $\Omega=\SI{100e-7}{\per\second}$ are rotating approximately $7
\times$ slower than Earth, and have a Rossby number $\Ro=\curly{O}(1)$), a retrograde moving forcing acts in a manner akin to further reducing the Rossby number of the flow, producing stronger \emph{effective} rotational regime.
This is observed in the inhibition of the superrotating jets produced in $\Ro=\curly{O}(1)$ planet, Figure~\ref{fig:zonal_jets} shows an extended cross-section for varying substellar velocity at rotation rate $\Omega=\SI{100e-7}{\per\second}$ where, reading from right-to-left an increasing retrograde substellar velocity turns off the superrotating equatorial jet, and turns on mid-latitudinal jets.
At the highest substellar velocities, both prograde and retrograde, $\tau_\C{diurnal} \ll \trad$ and the local equilibrium temperature is oscillating as the diurnal cycle passes over at a much quicker rate than the rate of thermal adjustment.
The consequence is a ``smearing out'' of the equilibrium profile, in the limit of infinitely fast substellar progression the relaxation profile at all longitudes would be the zonal mean of~(\ref{eq:teq}), shown in Figure~\ref{fig:mean_teq}.
Therefore on the planets with fastest moving substellar points, the forcing felt becomes more zonally symmetric, but also weaker.
At the surface the mean equator-to-pole temperature gradient is less than half that found at the substellar point, reducing the strength of subtropical zonal jets as expected from the thermal wind relation for the geostrophic component of the zonal wind \citep[\S 2.8.4, for example]{Vallis2017},
\begin{equation}
	f\dd{u}{p} = \frac{R}{p}\dd{T}{y},
\end{equation}
the magnitude of vertical shear in the zonal winds is proportional to the latitudinal temperature gradient.
% \begin{equation}
% 	\lim_{|s|\to\infty} T_{eq} = \frac{1}{2\pi}\int_0^{2\pi} T_{eq}
% \end{equation}


In general the propensity is for eastward flow in both the day and night hemispheres, either in an equatorial superrotating flow, or as mid-latitude jet structures, the tidally locked response in agreement with the previous work of~\cite{Edson2011},~\cite{Merlis2010} and~\cite{Noda2017}.
However, at low rotation rate and for intermediate retrograde substellar velocities $ 100 < s < \SI{0}{\meter\per\second}$ a \emph{westward} equatorial jet is generated.

\begin{figure}[tb]
	\centering
	\includegraphics[width=0.7\textwidth]{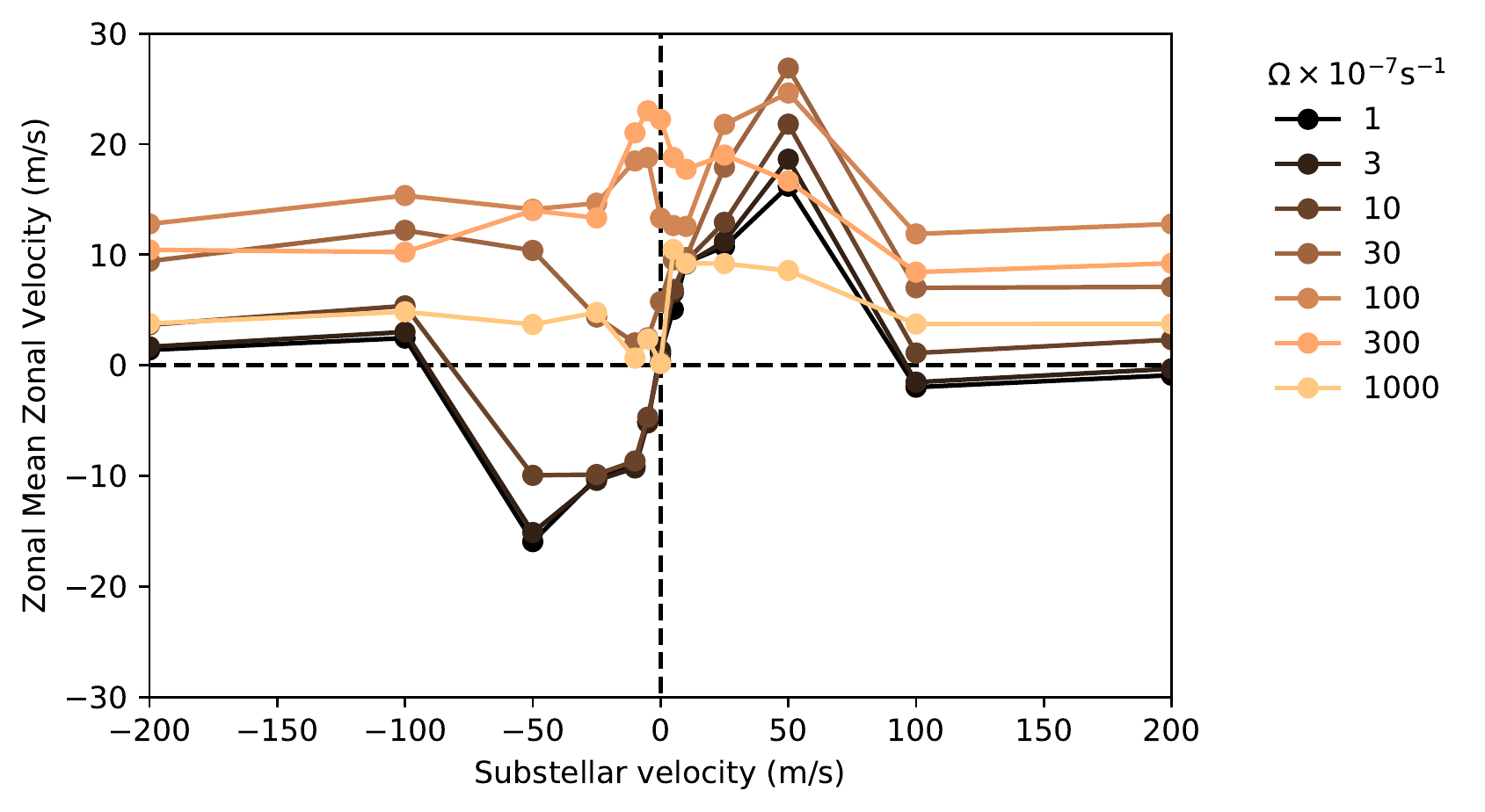}
	\caption{Zonal-mean zonal-wind in the upper troposphere (p=\SI{319}{hPa}) as a function of the substellar velocity at the equator.  Negative substellar velocity is retrograde, relative to the rotation of the planet.  As a point of reference, at equinox Earth would lie at \(\sim\SI{450}{m s^{-1}}\). Lines show increasing planetary rotation rate.  The transverse section view of this data, for three substellar velocities is shown in Figure~\ref{fig:zmzw_319_om_equator}.}\label{fig:zmzw_319_equator}
\end{figure}

Figure~\ref{fig:zmzw_319_equator} summarises the above, showing the zonal-mean zonal velocity in the upper troposphere as a function of substellar velocity (abscissa) and rotation rate (line) for the complete parameter space listed in Table~\ref{table:params}.
% When the planet is quickly rotating, geostrophic forces dominate, thermal wind balance and a strong superrotating jet at the equator dominate (compare to the longitude-latitude maps of zonal wind shown in Figure~\ref{fig:zonal_wind_319hPa}) and for all substellar velocities mean flow at this altitude is prograde relative to planetary rotation.
There appears to be a resonant response of the atmosphere at $|s| = \SI{50}{\metre\per\second}$ producing the strongest equatorial jets in either direction.
And this is the only point in the parameter space for which a significant equatorial subrotating flow is generated --- high Rossby number and retrograde substellar progression, $-50 \le s < 0$.
As the planetary Rossby number becomes very small ($\Omega \ge \SI{100}{\per\second}$), the upper level zonal flow becomes largely independent of diurnal cycle and substellar velocity, sphere at \(|s| \simeq \SI{50}{\meter\per\second}\), the zonal mean zonal velocity at height becomes a function of the  motion of the forcing.

% The gravity wave speed of a constantly stratified atmosphere can be related to linear shallow water theory (for example, Chapter~3~of~\cite{Vallis2017}); the shallow water wave speed is
% \begin{equation}\label{eqn:grav_speed}
%     c = \sqrt{g H_e},
% \end{equation}
% where $H_e$ is the ``equivalent depth'' shallow water layer.
% The stratified atmosphere can modally decomposed into a system of shallow water layers with equivalent depths
% \begin{equation}
% 	H_{e, n} = \frac{N^2}{g n},
% \end{equation}
% where $n$ is the eigenvalue of the vertical mode and $N^2$ is the square Brunt-Väisälä frequency
% \begin{equation}
% 	N^2 = \frac{g}{\tilde\theta}\dd{\tilde\theta}{z},
% \end{equation}
% for a mean potential temperature profile $\tilde\theta(z)$.
It can be shown that in the hydrostatic Boussinesq approximation (i.e. constant vertical stratification), the horizontal phase speed of wave propagation in a dry atmosphere is given by
\begin{equation}
	c_m = \frac{\omega}{\kappa} = \frac{N}{m},
\end{equation}
where $\kappa = k^2 + l^2$ is the horizontal wavevector, $m$ the vertical wavenumber \citep[\S 7.3]{Vallis2017}.
As $m \equiv \pi/H_c$, where $H_c$ the vertical wavelength, for the $m=1$ mode $H_c$ is twice the height of the troposphere, $H$, and so the fastest wave will be given by
\begin{equation}
	c_1 = \frac{NH}{\pi}. \label{eq:max_c}
\end{equation}

The Brunt-Väisälä frequency is defined as,
\begin{equation}
	N^2 \equiv \frac{g}{\tilde\theta}\dd{\tilde\theta}{z},
\end{equation}
for a mean potential temperature profile $\tilde\theta(z)$.
Near the surface, is $N^2$ largely constrained by the short relaxation times and therefore by the equilibrium forcing profile, but aloft is will be determined dynamically.
As a point of reference for this model, the largest ($m=1$) and fastest moving mode has been estimated from measurements for the Earth's troposphere, yielding a value $c\simeq\SIrange{44}{53}{\meter\per\second}$ for a deep tropospheric convection in response to moving heat forcing, given a free-tropospheric scale height $H=\SI{14}{km}$ \citep{Kiladis2009a}.

\begin{figure}[tb]
	\centering
	\includegraphics[width=0.8\textwidth]{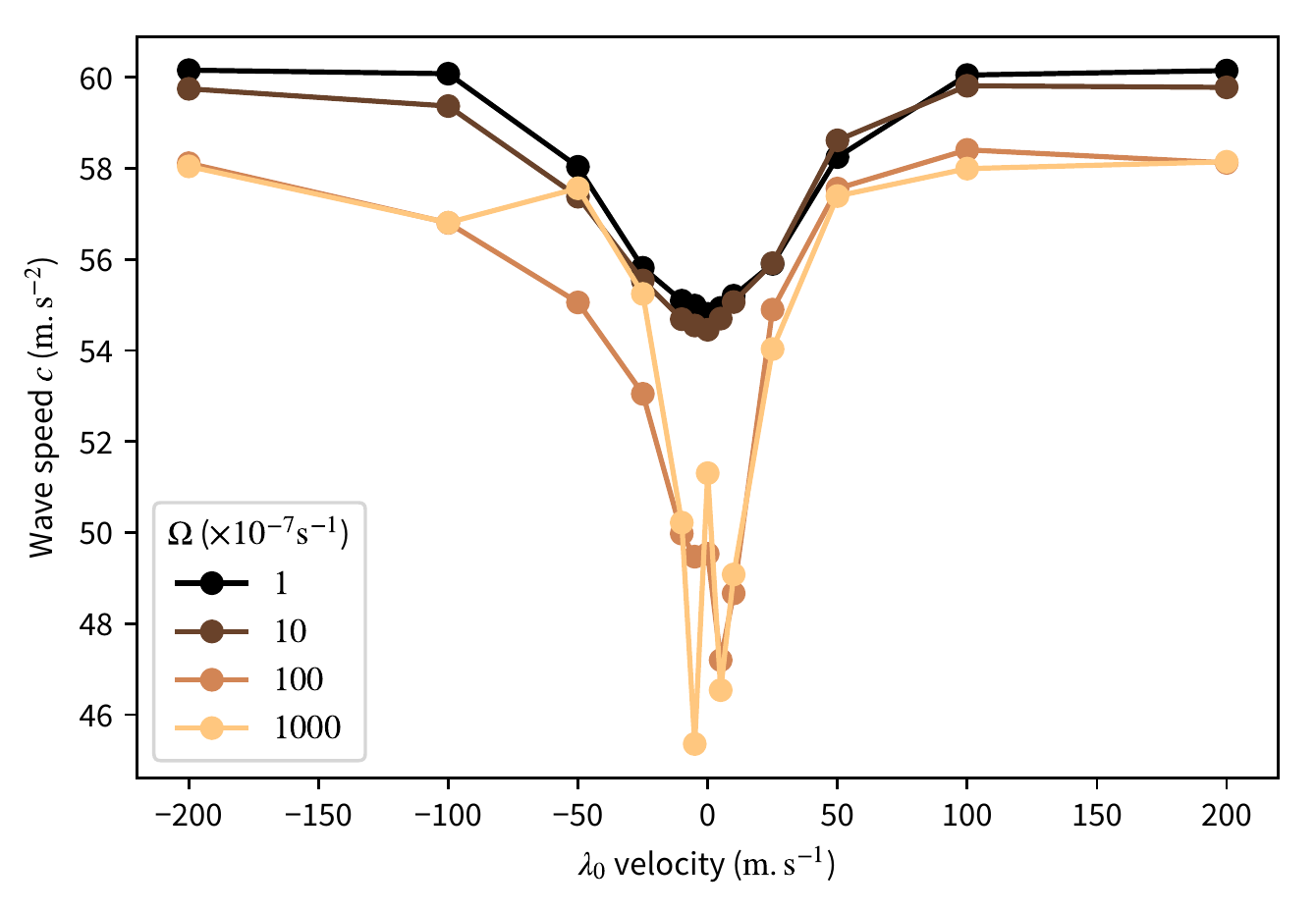}
	\caption{Phase speed of the fastest horizontal mode, $c=NH/\pi$ as a function of substellar velocity (abscissa) and increasing rotation rate (lines).
	}
	\label{fig:bv}
\end{figure}

We calculate a zonal-mean Brunt-Väisälä frequency and a scale height of the first baroclinic mode, $m=1$, in the equatorial free troposphere and find that it varies with the substellar velocity (Figure~\ref{fig:bv}), giving local gravity wave speeds between \SIrange{40}{60}{\metre\per\second} in the free troposphere, similar to those of Earth conditions, as we should expect from the essentially Earth-like parameters of the Newtonian cooling model.
This suggests that the largest response in the jet structure in the model at substellar velocities moving in either direction at $s \simeq \pm \SI{50}{\metre\per\second}$ may be a resonant interaction with the primary gravity wave mode.

\cite{Lindzen1981} demonstrates a theory of internal gravity waves interacting with the mean zonal wind;
internal gravity waves moving with a horizontal phase speed $c$ propagate vertically until they reach a critical level where $\bar u \simeq c$ and the waves break, depositing momentum.
Lindzen considered the interaction of gravity waves forced by the Earth's diurnal tide which has a phase speed of $\sim \SI{450}{\metre\per\second}$, much greater than the winds observed in the troposphere.
However, in our model of non-tidally locked exoplanets we are in a range where $|s|\simeq|\bar u|$ and thus a constructive or destructive deposition of zonal momentum can occur in the troposphere, with a peak response at the maximum internal gravity wave speed of the stratified atmosphere given by (\ref{eq:max_c}).
As frictional forces weaken with height and in general $\partial \bar u/\partial z > 0$, we would expect to see the height of maximum momentum deposition, and this is observed in Figure~\ref{fig:zonal_jets} --- the altitude of peak zonal velocity increases with substellar velocity.
As to why an overturning circulation is induced aloft (see the large vertical shear above the equator in the $s=\pm \SI{50}{\metre\per\second}$, $\Omega = \SI{10e-7}{\per\second}$ panels in Figure~\ref{fig:ss_as_zonal_wind}) requires further study.

\begin{figure}[tb]
	\centering
	\includegraphics[width=0.7\textwidth]{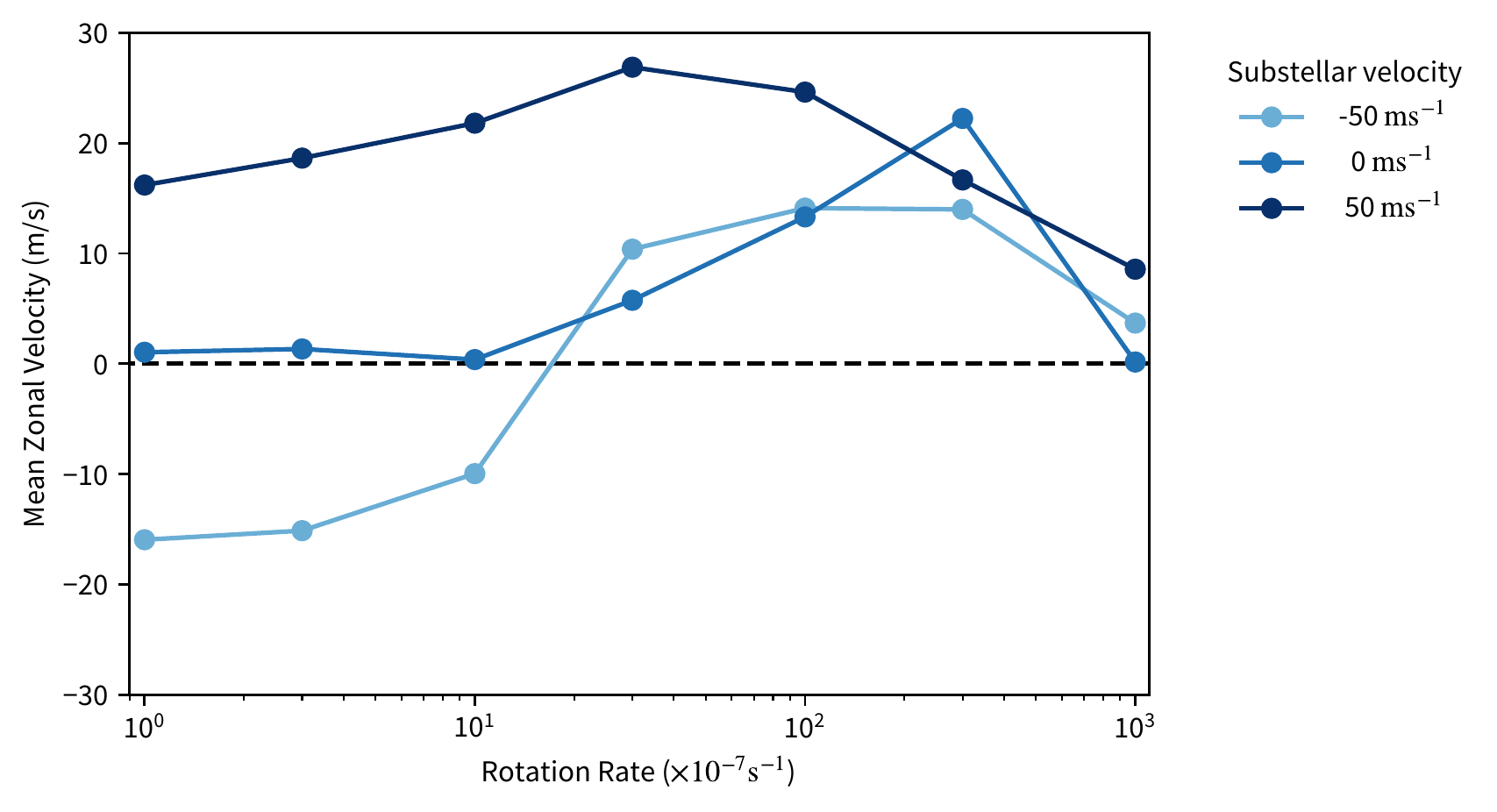}
	\caption{Globally-integrated mean zonal wind in the upper troposphere (p=\SI{319}{hPa}) as a function of rotation rate. Lines show the dependence on rotation rate at three substellar velocities: -50, +50, and \SI{0}{\meter\per\second}, corresponding to the local maxima/minima in seen in Figure~\ref{fig:zmzw_319_equator}.}\label{fig:zmzw_319_om_equator}
\end{figure}

It's clear that the horizontal circulation has a dependence on diurnal cycle at low rotation rates, if we were observing these planets from afar the advective heat transport associated with the flow may be detected as either an eastward or westward offset in the thermal phase curve.
In general, the upper-tropospheric mean zonal flow is additively modulated by the moving forcing, especially in high Rossby number regimes where the motion induced flow is significantly larger than the relatively weak overturning  flow of the tidally-locked case (Figure~\ref{fig:zmzw_319_om_equator}).

\begin{figure}[tbh]
\centering
\includegraphics[width=0.6\textwidth]{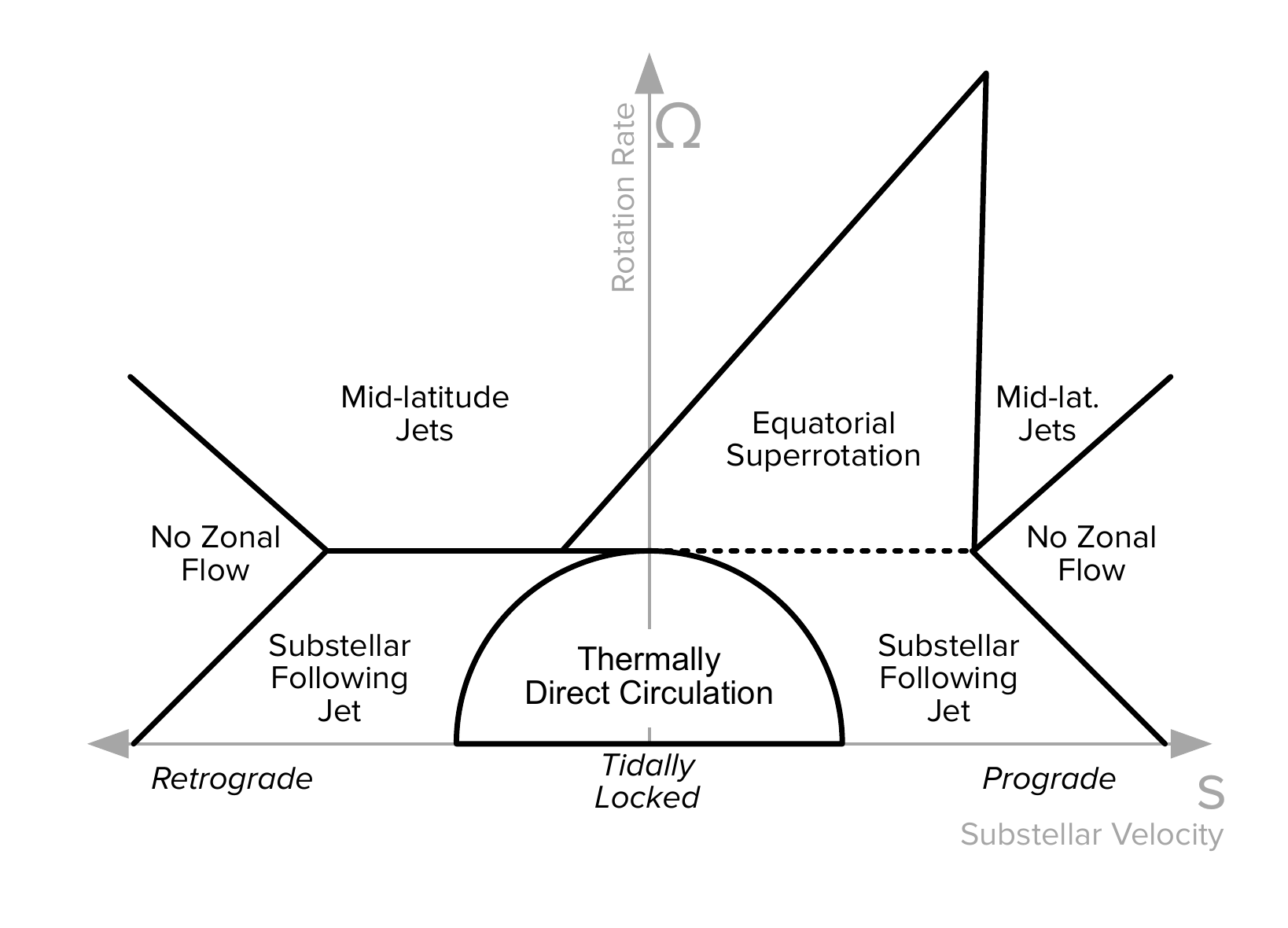}
\caption{The qualitative regime of the dynamics can be categorised by the nature of the upper-tropospheric flow.  This regime diagram classifies the circulation patterns observed across the parameter space of varying planetary rotation rate $\Omega$ and substellar velocity $s$.}\label{fig:ucomp_regime}
\end{figure}

Figure~\ref{fig:ucomp_regime} is a schematic overview of the dynamical regimes observed and presented in figures~\ref{fig:zonal_wind_319hPa},~\ref{fig:ss_as_zonal_wind}~and~\ref{fig:zmzw_319_equator}.
The region in which equatorial superrotation occurs is limited to a subset of rotational velocities, (clearest in the central column of Figure~\ref{fig:ss_as_zonal_wind}); very slowly rotating planets have a direct circulation in both directions around the planet, akin to two large Hadley--Walker cell circulations extending from substellar to antistellar points, and pole to pole.

The diversity of climate regimes at a given rotation rate can largely be understood as a result of the scale of timescales of radiative cooling, advection and the diurnal cycle.
In general, if $\trad \ll \tau_\C{adv, diurnal}$ the radiative forcing dominates and the thermal structure of the atmosphere will largely resemble the forcing, with significant zonal temperature gradients.

The redistributive effects of advection and the diurnal cycle act to reduce zonal variation, as seen in the upper rows of Figure~\ref{fig:lat_lon_temp_trop} where divergent advection from the substellar point efficiently redistributes heat globally.

\subsection{Thermal Phase Curves}\label{sub:phase_curves}

% Figure~\ref{fig:lat_lon_temp_surf}~and~\ref{fig:lat_lon_temp_trop} show the horizontal temperature distribution at slices at the near surface and in the mid-troposphere respectively.

% When the planet is slowly rotating and tidally locked we see large scale circulation over the entire day \& night side --- large scale divergence at height on the day side redistributes heat to the night side in a global Hadley cell pattern.

% At high rotation rate (lower rows of Figure~\ref{fig:lat_lon_temp_surf}~and~\ref{fig:lat_lon_temp_trop})

In our previous study using a shallow water model of the first baroclinic mode of the atmosphere \citep{Penn2017}, we demonstrated that both eastward and westward offsets in the thermal phase curve of a transiting exoplanet could be observed; the offset is sensitive to both absolute planetary rotation rate and the velocity of the substellar point.
It was shown that when the normalised substellar velocity $|s/c| < 1$, where $c=\sqrt{gH_e}$ is the linear shallow water gravity wave speed for a fluid of equivalent depth $H_e$, a dynamical balance could be maintained between a horizontal thermal gradient and the moving forcing.
When a theoretical ``thermal phase curve'' of the planet was calculated by integrating the height of the shallow water layer it exhibited offsets in both zonal directions relative to the substellar point.

In the same manner we again now calculate phase curves from the stratified model of the atmosphere.
Our model, with a prescribed relaxation temperature rather than diagnostic radiative heating, does not have an optical depth, nor does it need to satisfy an energy balance by radiating to space.
Therefore, there is no well-defined diagnostic outgoing longwave radiation (OLR), we use an approximation to this to calculate a thermal phase curve.
\begin{figure}[tbh]
\centering
\includegraphics[width=\textwidth]{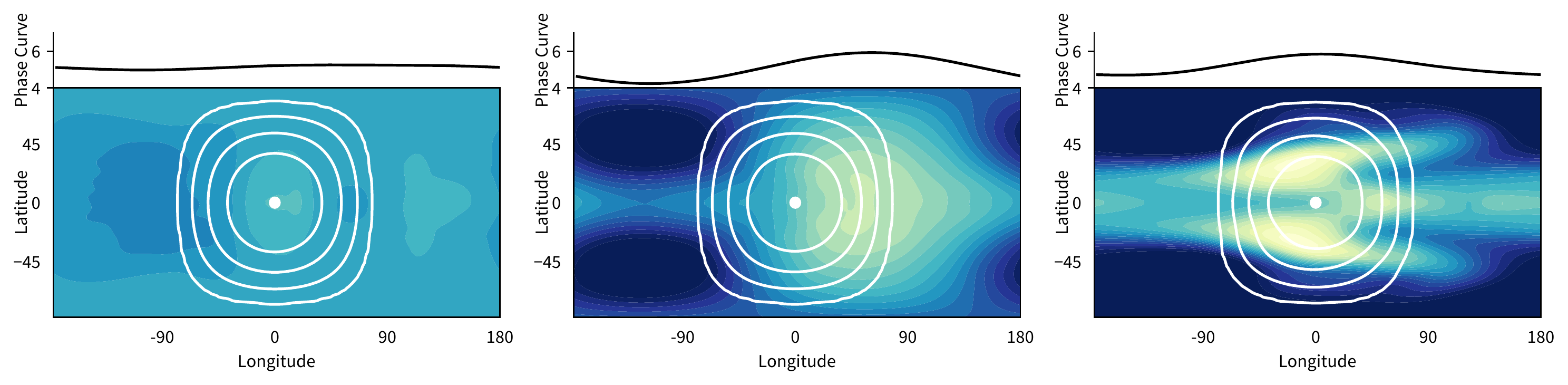}
\caption{Example temperature fields and their associated normalised phase curves.  From left-to-right, these correspond to $\Omega$ = 10, 100, \SI{1000e-7}{\per\second} in the central, $s=\SI{0}{\meter\per\second}$, column of Figure~\ref{fig:lat_lon_temp_trop}.
The brightest colour contour is \SI{260}{K}, with further contours decreasing in \SI{2}{K} intervals.  White contours show the position and extent of the relaxation temperature profile $T_{eq}$, the maximum of this at the substellar point indicated with a white dot. The phase curve \emph{offset} is given by the longitudinal separation between the substellar point and the peak of the phase curve, most clearly observed as a westward offset in the middle panel.}\label{fig:pc_examples}
\end{figure}
% The temperature at the top level of the model is not a good choice for emission temperature, stratosphere of the atmosphere is largely transparent to infrared radiation.
% It is also, typically, much colder than the theoretical emission temperature necessary for the planet to remain in energy equilibrium with stellar forcing.
%TODO: justification for choosing emission height in troposphere
%Following \cite{Vallis2017}, it can be shown that the atmospheric emission temperature, \(T_{eff}\), under a semi-gray approximation, is proportional to the temperature at the tropopause,\(T_{trop}\),
%\begin{equation}
%  T_{eff} = \frac{T_{trop}}{2^{1/4}},
%\end{equation}
%the factor of \(2^{1/4}\) coming
%as the height of the tropopause, the height where the atmosphere transitions from a convective-radiative equilibrium to isothermal radiative equilibrium.

We choose a radiating pressure level, \( p_{rad} \), as the height of emission to space.
At this level we can create a phase curve by performing a hemispheric integral of black body emission,
\begin{equation} \label{eq:phase_curve}
	I(\delta; p_{rad}) = \int_{\delta - \pi/2}^{\delta + \pi/2} \int_{-\pi/2}^{\pi/2} a^2 \sigma T^4(\lambda, \phi, p_{rad}) \cos\lambda \cos^2\phi \d\phi  \d\lambda,
\end{equation}
where \( \delta \) is the observational zenith longitude,
% and \( \sigma = \SI{5.67e-8}{W m^{-2} K^{-4} } \) is the Stephan-Boltzmann constant.
and $\sigma$ is the Stephan-Boltzmann constant.
The \( \cos \lambda \cos \phi \) factor comes from the projection of the curved surface of the planet onto a flat observational disc --- the emission received is proportional to the distance from the centre of the disc.

The phase curve is normalised by the hemispheric integral of the night-side of the planet equilibrium temperature, \(T_{strat}\), and by moving into the reference frame of the moving substellar forcing,
\begin{equation} \label{eq:phase_curve_normed}
	\hat I(\delta; p_{rad}) = \int_{\delta - \pi/2}^{\delta + \pi/2} \int_{-\pi/2}^{\pi/2} {\left( \frac{T(\xi, \phi, p_{rad})}{T_{strat}} \right)}^4 \cos\xi \cos^2\phi \d\phi  \d\xi,
\end{equation}
where \( \xi = \lambda - \lambda_0(t) \) and the observational zenith longitude, \( \delta \), is now relative to the substellar point (see
Figure~\ref{fig:pc_examples} for examples of the temperature field and synthetic thermal phase curve for tidally locked exoplanets with varying rotation rate).

\begin{sidewaysfigure}[tb]
	\centering
	\includegraphics[width=\textwidth]{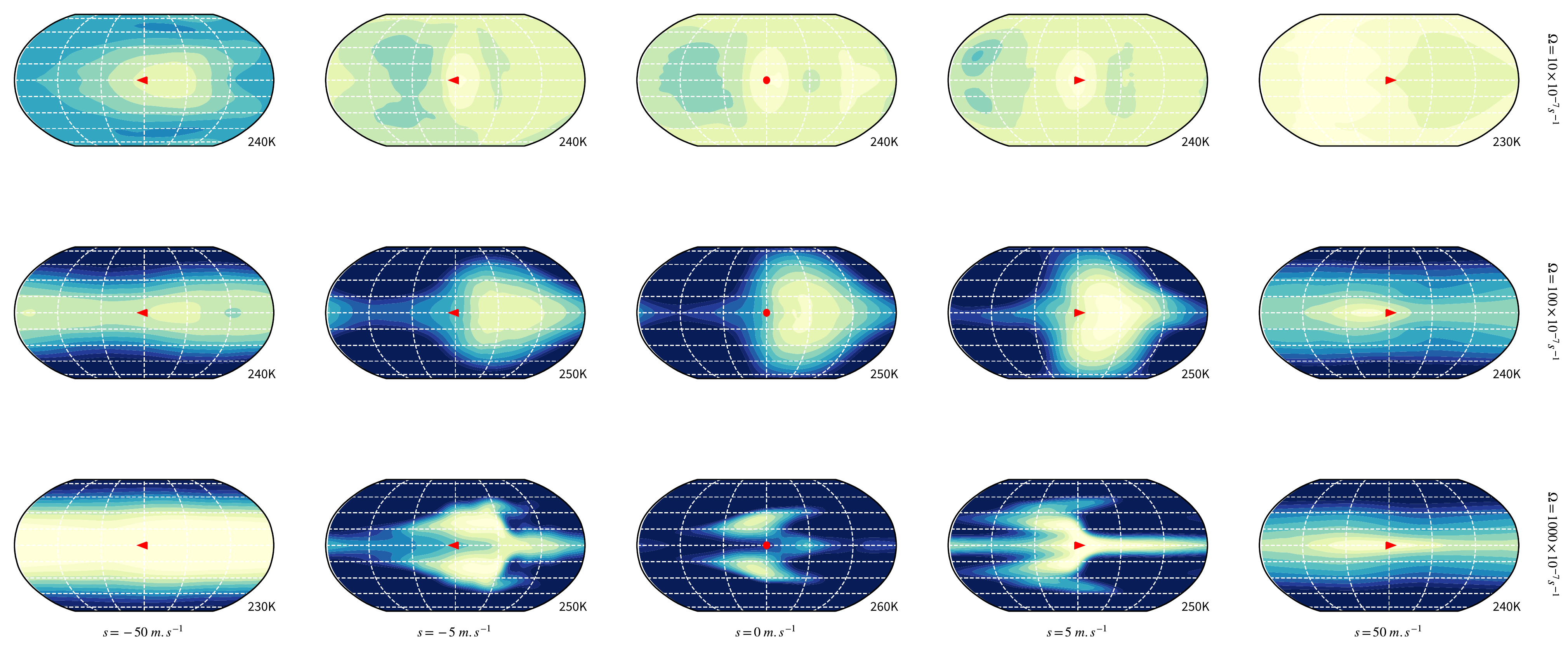}
	\caption{Mid-tropospheric atmospheric temperature (p = \SI{658}{hPa}).   Temperature reference in the lower right corner of each pane indicates the value of the highest temperature contour, with successive cooler contours at \SI{2}{K} intervals.  The central column (tidally locked) phase curve examples are shown in Figure~\ref{fig:pc_examples}. At this height in the atmosphere $\trad=\SI{20}{days}$, there is no frictional damping.}\label{fig:lat_lon_temp_trop}
\end{sidewaysfigure}

We first consider the thermal response in three dimensions before addressing the contraction of this into a one dimensional phase curve.
Figures~\ref{fig:lat_lon_temp_trop} presents the atmospheric temperature in  mid-troposphere, just above the prescribed frictional boundary layer, for varying rotation rate and substellar velocity.

% \begin{sidewaysfigure}[tb]
% 	\centering
% 	\includegraphics[width=\textwidth]{lat_lon_temp_P943_varying_5_reduced}
% 	\caption{Near-surface boundary layer atmospheric temperature.  Temperature reference in the lower right corner of each pane indicates the value of the highest temperature contour, with sucessive cooler contours at \SI{5}{K} intervals.  At this height in the atmosphere $\trad=\SI{5}{days}$, $\tau_\C{fric} = \SI{1}{day}$.}\label{fig:lat_lon_temp_surf}
% \end{sidewaysfigure}
In the stratified model we find a more complicated relationship than in the shallow water study.
At the surface, in the  the thermal response is tightly coupled to the forcing, and at all rotation rates the peak integrated temperature is lagging the motion of the substellar point.
However at altitude, from where observed infrared emission will originate, the temperature distribution is sensitive to the speed of diurnal cycle.

Above the boundary layer momentum and temperature are less strongly damped, thermal advection is efficient compared to friction and the hotspot is advected by the wind field.
Figure~\ref{fig:lat_lon_temp_trop} shows the horizontal temperature structure mid-troposphere; for both eastward and westward propagating substellar points eastward and westward hotspots are observed, depending on the rotation rate of the planet.

As previously demonstrated in the hot Jupiter studies of~\cite{Showman2015}, a quickly moving substellar point, or rapidly rotating system, results in a reduction in zonal variability.
At the mid-troposphere example plots shown in Figure~\ref{fig:lat_lon_temp_trop} there is a $\sim\SI{65}{K}$ gradient between substellar and anti-stellar point in the equilibrium temperature $T_{eq}$; in contrast the the atmospheric temperature varies little, an efficient thermal transport from day to night side.
The efficient heat transport and zonal redistribution with a rapidly moving substellar velocity predicts relatively warm night-side temperatures and a largely isothermal distribution of temperature in the upper atmosphere, this is perhaps expected from our intuition of Earth, which has a very rapid diurnal cycle relative to atmospheric radiative timescales, $\tau_{\mathrm{diurnal}} \ll \tau_\mathrm{rad}$.
The reduction in the meridional temperature gradient is also a result of the short diurnal timescale, providing an effective relaxation profile as shown in Figure~\ref{fig:mean_teq} and discussed above.

\begin{figure}
\gridline{
	\fig{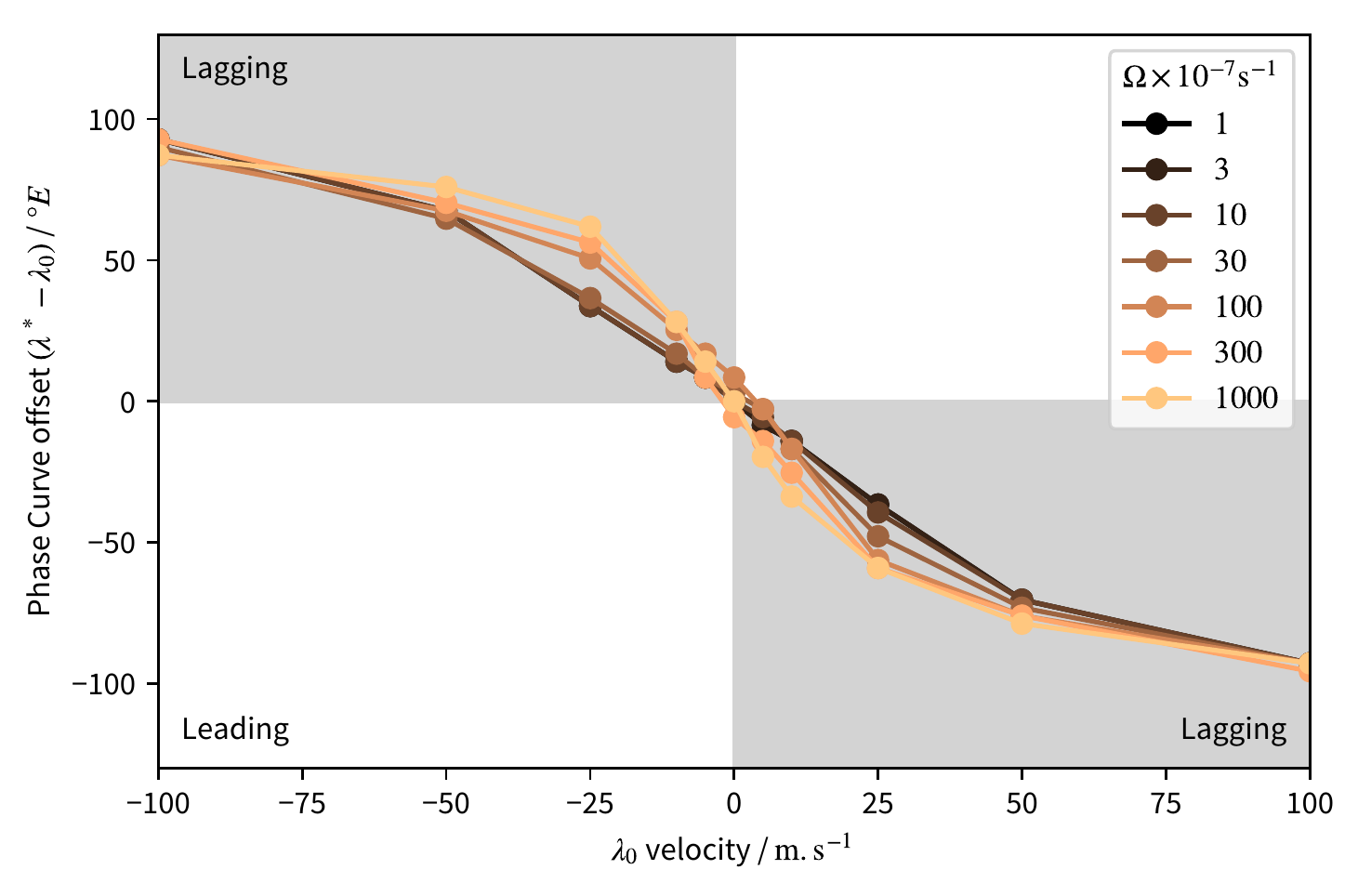}{0.4\textwidth}{(a) \SI{943}{hPa}}
	\fig{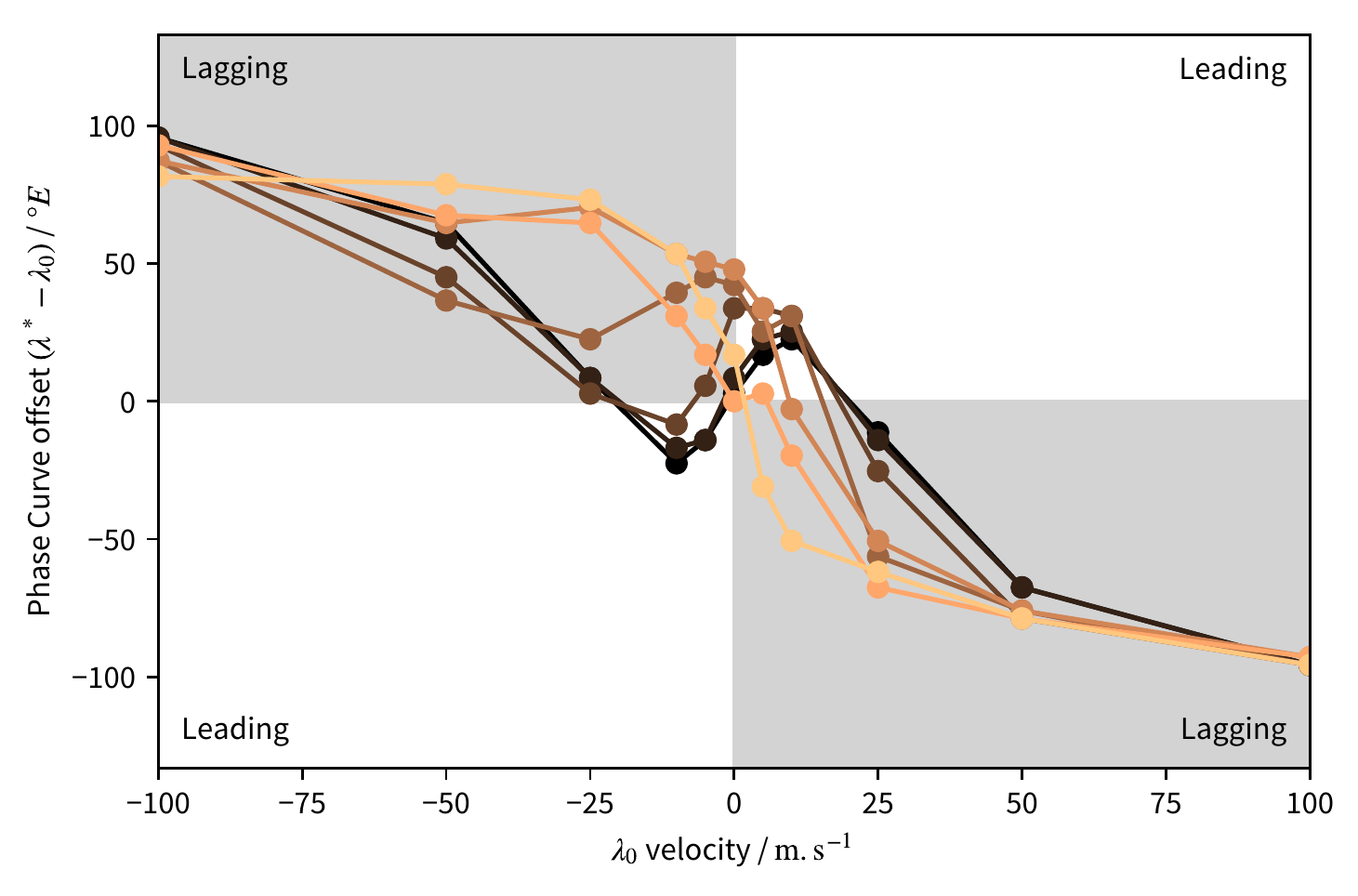}{0.4\textwidth}{(b) \SI{742}{hPa}}
}
\gridline{
	\fig{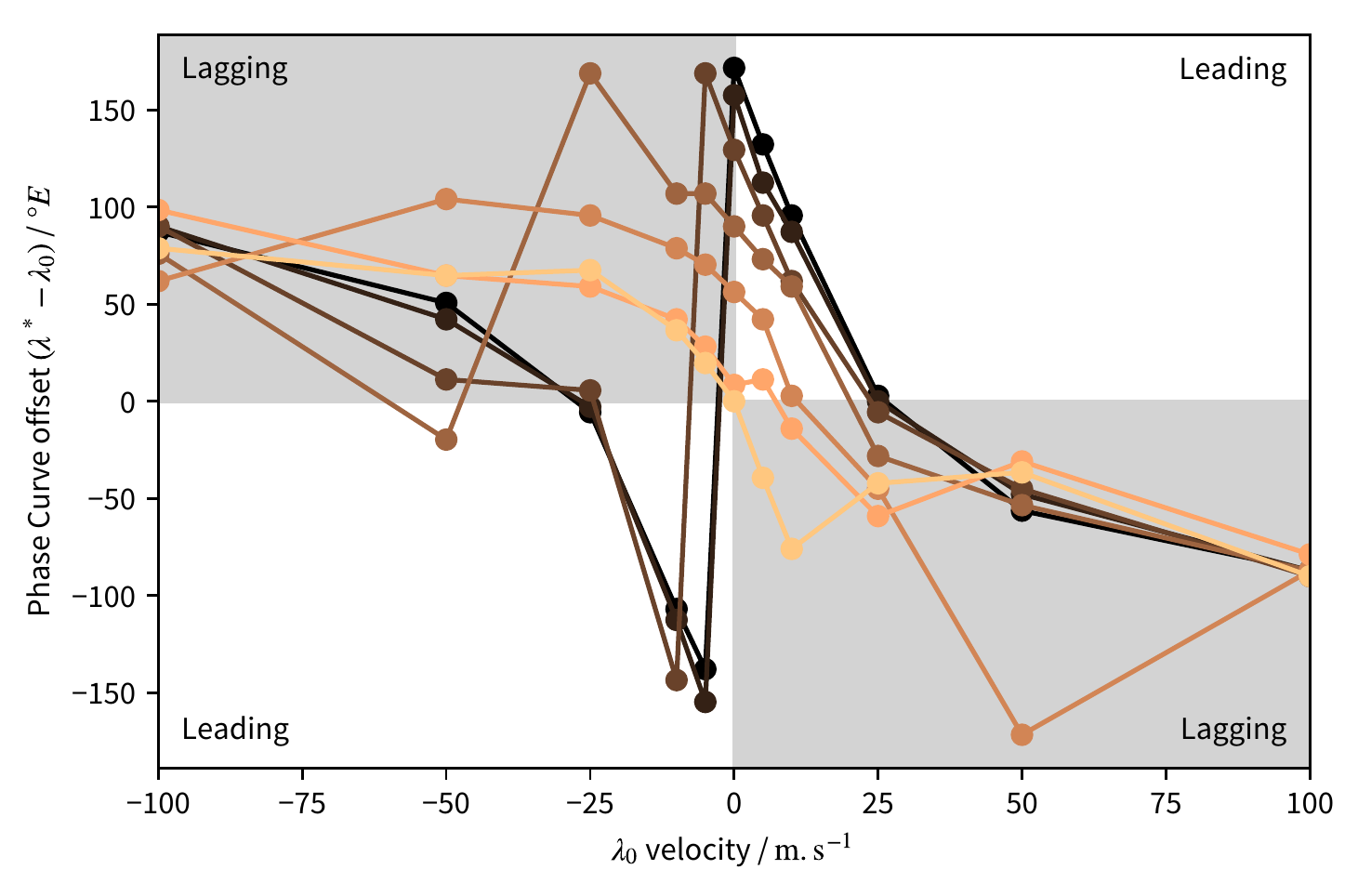}{0.4\textwidth}{(c) \SI{583}{hPa}}
	\fig{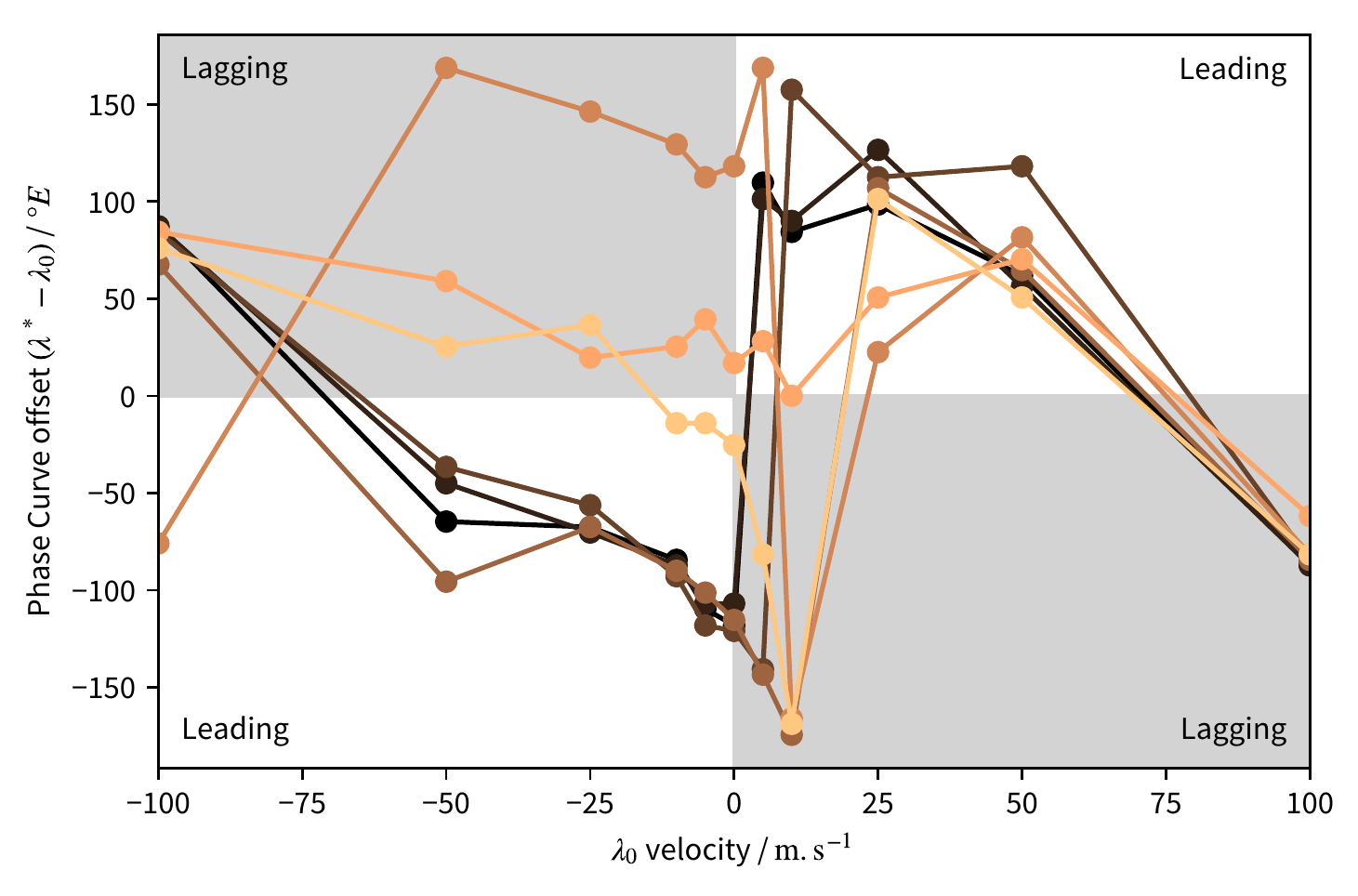}{0.4\textwidth}{(d) \SI{407}{hPa}}
}
\caption{Phase curve offsets for thermal emission from different heights in the atmosphere, as a function of substellar velocity velocity $s$ (abscissa).  Lines of increasing brightness show the response at increasing planetary rotation rate.  The segments where the thermal hotspot leads the progression of the substellar point (e.g. eastward substellar motion, eastward hotspot offset) are shaded white, the lagging segments shaded grey.} \label{fig:phase_curves}
\end{figure}
\begin{figure}[tb]
	\centering
	\includegraphics[width=0.4\textwidth]{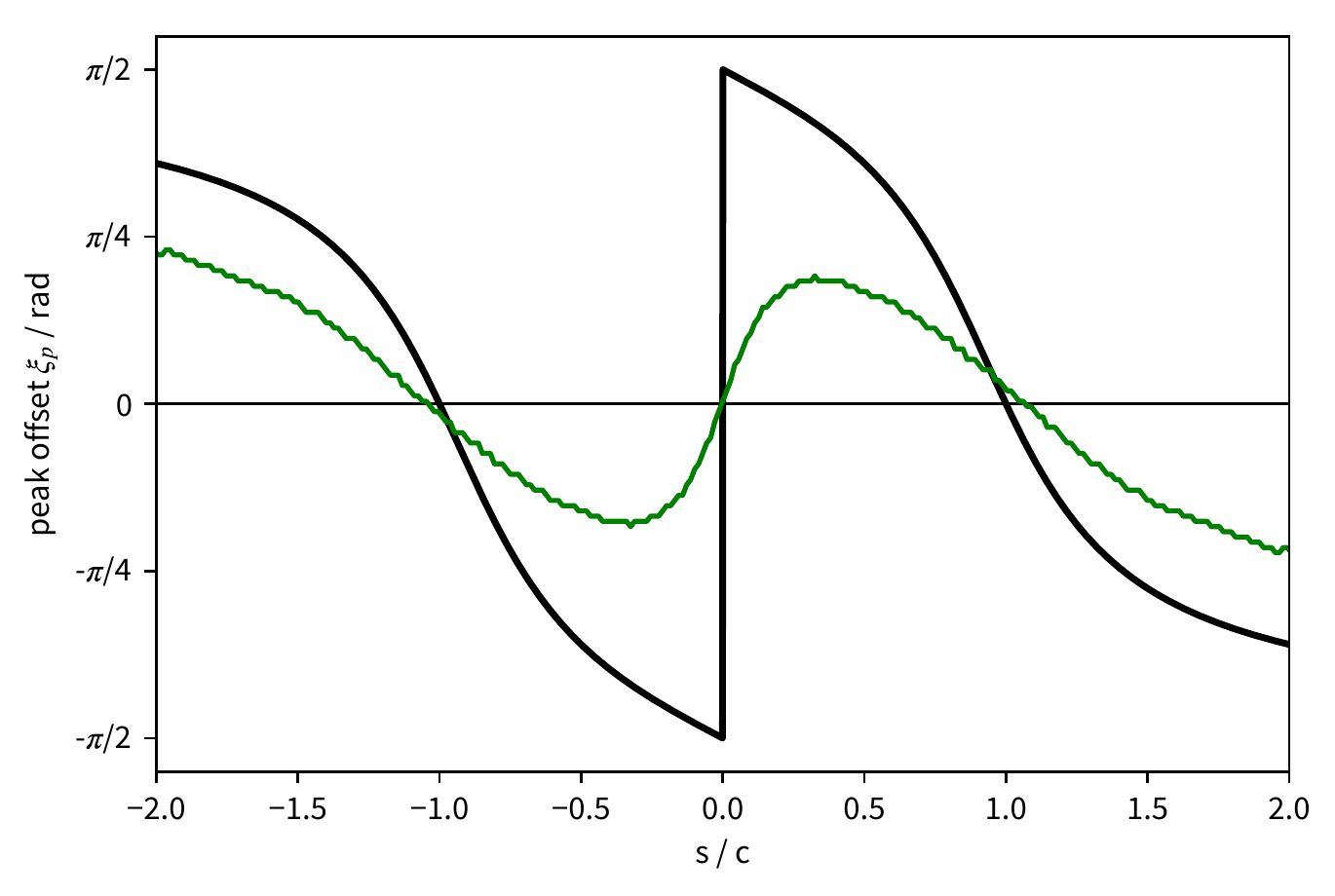}
	\caption{Analytic phase curve offset for an analogous one-dimensional shallow-water model along the equator in the absence of rotation.  The black line is (\ref{eq:xi_p}), the green line shows the results of a numerical simulation that includes non-linear terms, momentum drag and a non-smooth forcing similar in form to (\ref{eq:teq}).}\label{fig:analytic_equator}
\end{figure}

The phase curve \emph{offset} is found by calculating the normalised phase curve relative to the substellar point using (\ref{eq:phase_curve_normed}), and then solving for the longitude of maximum integrated thermal emission.
The panels of Figure~\ref{fig:phase_curves} show the phase curve offset as a function of substellar velocity near the surface and at several levels in the troposphere.
Phase curves near the surface are a strong function of substellar velocity, with little influence from rotation rate; in general the faster the substellar point moves, the further lagged the thermal hotspot is behind the point of maximal heating.
This is due to the short timescales of both radiative and frictional damping in the bottom boundary of the model, wind velocity at this level is small resulting in little advection, and strong relaxation towards the heating profile quickly eliminates thermal inertia from the passing substellar point.
It was previously shown \citep{Penn2017} that in the simple case of a linear non-rotating one-dimensional shallow-water model heated periodically along the equator, the offset between forcing and response, $\xi_p$, is given by the form
\begin{equation}
	\xi_p = \arctan\left( \frac{(c^2 - s^2) \trad}{s} \right), \label{eq:xi_p}
\end{equation}
where $c$ is the shallow-water gravity wave speed, which tends to a finite limit of a half phase lagging offset between forcing and substellar point as the substellar velocity increases in either direction (see Figure~\ref{fig:analytic_equator}),
\begin{equation*}
	\lim_{s\to\mp\infty} \xi_p = \pm \pi/2.
\end{equation*}
However, it also has a maximum as we approach the tidally-locked,
\begin{equation*}
	\lim_{s\to\pm 0} \xi_p = \pm \pi/2,
\end{equation*}
a discontinuous \emph{leading} offset that becomes smooth through the tidally-locked state with the addition of non-linear effects.
This character of the response is retained in the stratified model, lagging responses tending to a $\pi/2$ limit in the case of fast substellar motion.
In the troposphere where frictional damping is not applied and the timescale of thermal relaxation is much longer (panels (c) and (d) of Figure~\ref{fig:phase_curves}), a leading offset up to $\SI{20}{\degree}$ is seen for both prograde and retrograde moving diurnal cycles on slowly rotating planets expected from the shallow water theory.
As $\Ro < 1$ the circulation becomes increasingly constrained to equatorial regions, equatorially trapped Rossby gyres are seen as temperature maxima above and below the equator, and a frictionally trapped Kelvin wave just east of the substellar point (most clearly shown, and their impact on the integrated thermal phase curve, in the last panel of Figure~\ref{fig:pc_examples}).

In the upper atmosphere all our Newtonian cooling models show a largely isothermal response in all directions, the magnitude of $\Delta T_\C{day-night}$ becomes negligible and so the offset becomes highly variable and dominated by synoptic changes.
% This is a limitation of the relaxation model used in this study, using a more realistic radiation parameterisation would yield a more complex thermal response above the deep-convective layer.

% The question of determining rotation rate from orbital phase curve has been addressed before (for example,~\cite{Rauscher2014, Cowan2011}).
The model presented here shows that it is possible to observe both an eastward and westward offset in the observed thermal phase curve of an exoplanet, with few assumptions made about the composition of the atmosphere, beyond the heat capacity and gravitational constants of Earth.
The simple forcing parameterisation offers both advantages and disadvantages in this respect.
The relaxation is linear in temperature, scaling the atmospheric height appropriately, we might expect the results to hold at hotter temperatures of some of the already observed close-in exoplanets.
The assumptions we make about the composition of the atmosphere determine the lapse rate of the equilibrium profile and the internal gravity wave speed, the effect of changing the shape of this vertical profile to that of e.g. a H/He atmosphere would be an interesting extension.

In the disadvantages column, the lack of a strict energy balance or a two-steam radiation scheme means that the thermal emission of the model can only be considered at a specific height within the atmospheric column
An observed multi-band phase curve will information from more than one height in the atmosphere and this can't easily be directly compared to a single height in the atmosphere \citep{Dobbs-Dixon2017}.
More complex models have addressed the hotspot of specific exoplanet candidates and find the offset can vary significantly depending on the radiating level of the atmosphere \citep{Hammond2017}.
An interesting further step would be to pass the temperature-pressure profiles of the results shown above to a radiation code to calculate a true thermal emission field.
This would require making further choices about the composition of the atmosphere that are beyond the scope of this idealized study.

\section{Conclusions}
We have used a Newtonian cooling / Rayleigh friction model of the atmosphere to simulate a suite of Earth-like exoplanets in the parameter regime around a state of tidal-locking.
We showed that the offset in the thermal phase curve is affected by both the rotation rate and diurnal cycle of the planet, with eastward and westward offsets from the substellar point observed at a range of parameter values.
Linking this to an \emph{observed} phase curve is not necessarily straightforward however, care must be taken when making further inference as the offset also has a strong dependence on the depth within the atmosphere from which the thermal emission originated.

Upper tropospheric zonal winds are sensitive to both the rotation rate and diurnal period of the planet; even a small amount of asynchronous rotation can induce large changes in the circulation, for example inhibiting the formation of the equatorial superrotating jet predicted by many tidally locked exoplanet models.
In the tidally locked state a robust superrotating equatorial jet is observed over a range of rotation rates typical of a planet in a 5--100 day orbit.
However with only a small deviation from the tidally locked state, producing a moving diurnal heating pattern, the superrotating equatorial jet can be split into mid-latitudinal jet streams.
The effect of the diurnal cycle on atmospheric dynamics is most pronounced when the substellar velocity has similar magnitude to the internal wavespeed of the stratified atmosphere, depositing momentum into jets in the same direction as the moving forcing either constructively strengthening the superrotation (in the case of a eastward moving forcing) or destructively weakening it (for westward moving substellar forcing).

This study demonstrated that asynchronous rotation, even when it is close to being synchronous, can be crucial in understanding the large-scale circulation of a planet's atmosphere.
As our methods of detection improve and we discover terrestrial planets further from their host star, with exomoons, and/or with thick Vesuvian-like atmospheres, the likelihood of asynchronous rotation increases, and with it a whole new range of interesting dynamical regimes to be explored.

% section conclusions (end)
	\bibliography{hs_exo_biblio}
\end{document}